\documentclass[traditabstract]{aa} 
\usepackage{natbib}
\usepackage{supertabular}
\usepackage{longtable}
\usepackage{graphicx}
\usepackage{lscape}
\usepackage{ulem}
\usepackage{multirow}
\usepackage{rotating}
\usepackage{dcolumn}
\newcolumntype{d}{D{.}{\cdot}{-1}}
\begin{document}

\title{CH in stellar atmospheres: an extensive linelist} 

\author{Masseron, T.$^1$
\and
Plez, B.$^2$
\and
Van Eck, S.$^1$
\and
Colin, R.$^3$
\and
Daoutidis, I.$^1$
\and
Godefroid, M.$^3$
\and
Coheur, P.-F.$^3$
\and
Bernath, P.$^4$
\and
Jorissen, A.$^1$
\and
Christlieb, N.$^5$
}
\institute{ Institut d'Astronomie et d'Astrophysique, Universit\'e Libre de
Bruxelles (ULB), CP 226, Boulevard du Triomphe, B-1050 Bruxelles, Belgium 
\and
Laboratoire Univers et Particules de Montpellier , Universit\'e Montpellier 2, CNRS, F-34095 Montpellier, France
\and
Service de Chimie quantique et Photophysique, Universit\'e Libre de
Bruxelles, CP160/09, avenue F.D. Roosevelt 50, 1050 Bruxelles
\and 
Department of Chemistry \& Biochemistry, Old Dominion University, Norfolk, VA, USA
\and
University of Heidelberg,
   Zentrum f\"ur Astronomie, Landessternwarte, K\"onigstuhl 12, 69117
   Heidelberg, Germany
}

\abstract{The advent of high-resolution spectrographs and detailed stellar atmosphere modelling has strengthened the need for accurate molecular data. Carbon-enhanced metal-poor (CEMP) stars spectra are interesting objects with which to study transitions from the CH molecule. We combine programs for spectral analysis of molecules and stellar-radiative transfer codes to build an extensive CH linelist, including predissociation broadening as well as newly identified levels. We show examples of strong predissociation CH lines 
in CEMP stars, and we stress the important role played by the CH features in the Bond-Neff feature depressing the spectra of barium stars by as much as 0.2~magnitudes in the $\lambda=$3000 -- 5500~\AA\ range. Because of the extreme thermodynamic conditions prevailing in stellar atmospheres (compared to the laboratory), molecular transitions with high energy levels can be observed. Stellar spectra can thus be used to constrain and improve molecular data. 
 }

\maketitle
\section{Introduction}
The computation of stellar atmosphere models requires, among various other ingredients, the knowledge of radiative transition rates for a large number of atomic and molecular species \citep{Gustafsson2008}. Extensive linelists, as complete as possible, with accurate oscillator strength values are needed for the calculation of opacities that affect the thermal structure of the atmosphere. The computation of detailed emergent spectra also requires accurate line positions \citep{2008PhST..133a4003P}.
The advent of high-resolution spectrographs on 8m-class telescopes (e.g. VLT/UVES, Keck/HIRES, VLT/CRIRES, SUBARU/HRS) allows a very detailed study of spectra of many astronomical objects. 
The discussion of possible abundance pattern differences between stars with and without planets lies now at a level of the order of 0.01~dex \citep{2010Ap&SS.328..185G,2005ARA&A..43..481A}.
The conclusions therefore heavily depend on the quality of the whole chain of data acquisition and analysis, not the least of which are the physical data. This kind of study requires a line position accuracy of the order of a few hundredths of \AA\ or cm$^{-1}$ in the optical domain, and oscillator strengths with an accuracy better than 5\%.

Great efforts have been devoted to improving atomic and molecular line data, e.g. VALD \citep{Kupka1999} and DREAM \citep{Biemont1999} for atoms, and \citet{Bernath2009} and EXOMOL \citep{Tennyson2012} for molecules. In particular, many transitions have been analysed in the laboratory, leading to a large number of accurate line positions, and molecular constants. However, line strength information is often missing, and lines involving highly excited states, not visible in the laboratory but important in warm astrophysical environments like stellar atmospheres, are lacking. We are therefore aiming at constructing molecular linelists for astrophysical use. The present paper targets CH.

The CH molecule is one of the most studied free radicals because it forms in various physical and astrophysical environments. This molecule was first detected by \citet{Heurlinger1919} and then studied by many scientists \citep[see references in][]{Kalemos1999}. This molecule has many valence and Rydberg states as reviewed by \citet{Vazquez2007}, but only the lower energy states are of interest for stellar astrophysics. Our aim in this paper is to improve the situation for the ro-vibrational X$^2\Pi$-X$^2\Pi$ and the A$^2\Delta$-X$^2\Pi$, B$^2\Sigma^-$-X$^2\Pi$, and C$^2\Sigma^+$-X$^2\Pi$ electronic transitions of CH (for both $^{12}$CH and $^{13}$CH isotopologues). 
Bound-bound transitions for CH appear in the near UV spectra of almost all F, G, K stars, whatever their carbon enrichment is. The G band at 4300~\AA\ is particularly prominent, and is used, for example as a criterion for stellar classification (CH stars), and as a proxy for magnetic flux concentrations in the solar photosphere \citep{2004A&A...427..335S}. As another example the $^{12}$C/$^{13}$C isotopic ratio can be derived from the CH molecular lines in the near UV, and are often the only available carbon abundance indicator in very metal-poor stars. In addition, a good knowledge of the CH lines in the blue-UV part of the spectrum is also a prerequisite for determining the abundance of heavier metals (e.g. Bi, Pb, and U) in stellar environments, as these lines often blend those of the atomic species.
To date, three linelists exist for CH. The first two have been available for a long time: one from R. Kurucz\footnote{Bob Kurucz' CD-ROM 15 or http://kurucz.harvard.edu/linelists.html}, for which there is little information on how the linelist was computed, and the SCAN-CH of \citet{Jorgensen1996}, which also contains the ro-vibrational X-X transition. More recently, we compiled a third linelist, published as a poster
at the ``14th Cool Stars, Stellar Systems, and the Sun'' conference in Pasadena, in 2006, and distributed since then by Plez.
This list was assembled after our discovery that unidentified strong and broad
lines in the near-UV spectrum of carbon-enhanced metal-poor (CEMP) stars were due to
predissociation lines of CH (see the Appendix).
The line positions and gf values come from the LIFBASE program \citep{LIFBASE1999};
excitation energies from the SCAN linelist, and isotopic shifts were recomputed by our team.

It is this last linelist that is presented and discussed here, but we have largely updated it by (i) the computation of a uniform set of molecular constants from laboratory data supplemented by our measurement of high excitation lines in stellar spectra, (ii) the inclusion of both the observed and calculated line positions, (iii) the inclusion of radiative damping parameters accounting for pre-dissociation, and (iv) the inclusion of the ro-vibrational X-X transition. The paper is structured as follows: first, we explain the method used to derive the linelist (Sect.~\ref{sec:methodo}), then we describe the data we have used for $^{12}$CH and $^{13}$CH (Sect.~\ref{sec:obs_trans}), and finally we summarize the content of the list (SecT.~\ref{sec:results}); we illustrate its improvements for stellar spectra in the Appendix.

%%%%%%%%%%%%%%%%%%%%%%%%%%%%%%%%%%%%%%%%%%%%%%%%%%%%%%%%%%%%%%%%%%%%%%%%%%%%%%%%%%%%

\section{Method}\label{sec:methodo}
At least two parameters are required to assemble a molecular linelist: line positions and absolute line intensities. Other parameters may be added, e.g. collisional broadening parameters or Land\'e factors. Ab initio calculations can be attempted, but the use of observed line positions and intensities is more reliable.
We therefore collected extensive laboratory data for $^{12}$CH and $^{13}$CH, which we supplemented with line positions extracted from stellar spectra. 

We followed the method outlined in \citet{Li2012}. 
First, we used the PGopher\footnote{PGOPHER version 8.0, a Program for Simulating Rotational Structure, C.M. Western, University of Bristol, http://pgopher.chem.bris.ac.uk/.} software to determine molecular constants from line  positions. For the specific case of CH, these molecular constants include vibrational, rotational, and spin-rotational constants for all electronic states as well as spin-orbit and $\Lambda$-doubling constants for the X$^2\Pi$ and A$^2\Delta$ states. The line positions used as input to PGopher were weighted according to the quoted accuracy of the various sources.

\begin{figure*}[h!]
\includegraphics[width=14cm,angle=-90]{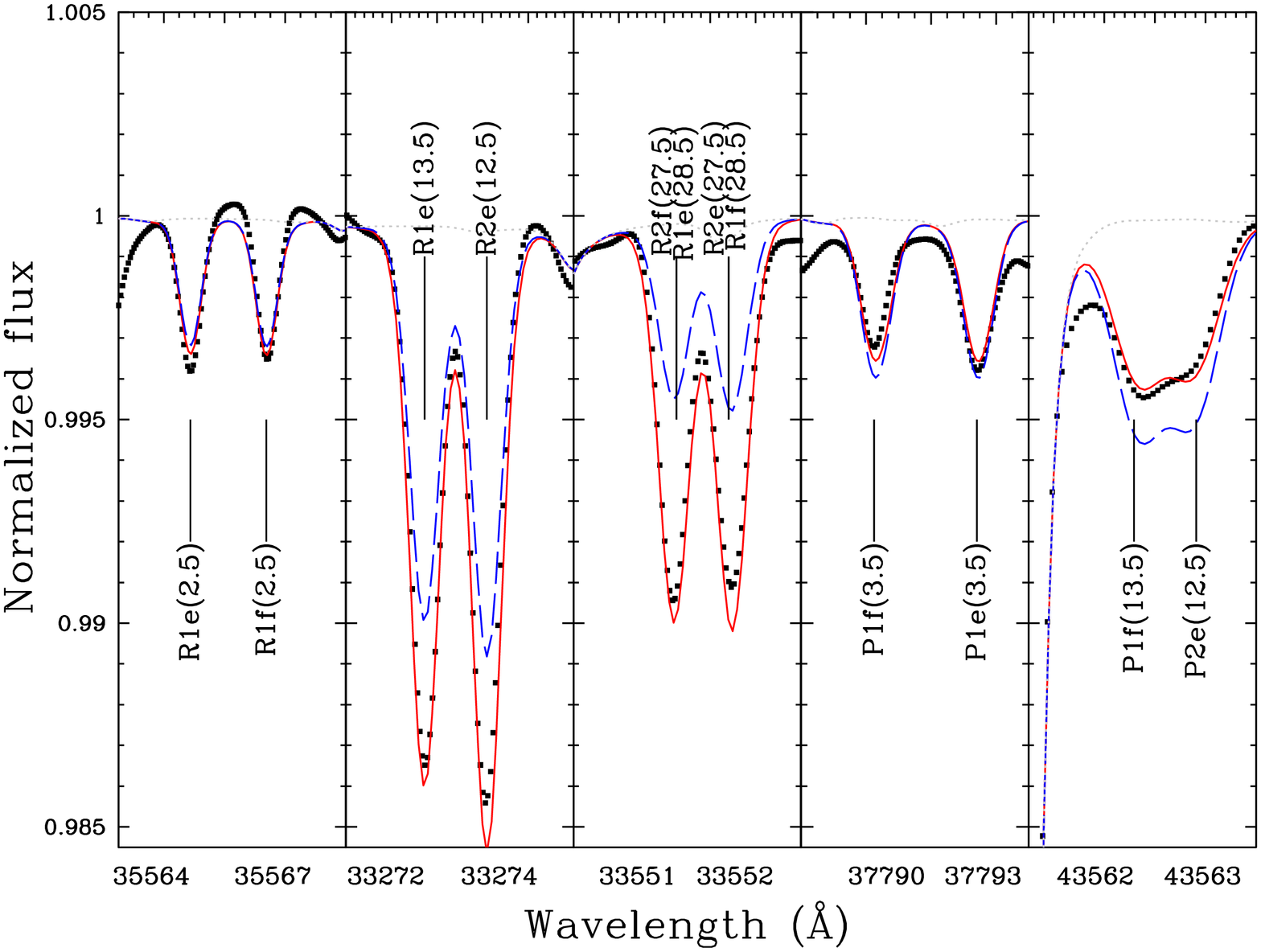}
\caption{Synthetic spectrum of the Sun near the 1-0 ro-vibrational band of CH compared to the observed solar spectrum \citep[ACE;][]{Bernath2005} (dotted line). The red line includes the J-dependence correction (see text) while the blue does not. The first 3 panels from left to right show from the R branch, for which intensity increases with increasing J. The increasing relative strength between the synthesis with and without the J-dependence correction demonstrates the rotational dependence of the transition moment for CH. The two right  panels show lines from the P branch. The fact that the J-dependence correction is not the same as for R branches is due to the Herman-Wallis effect.}
\label{fig:rot_depend}
\end{figure*}

The version of PGopher we use is able to generate linelists and also provides H\"onl-London factors for doublet states. The computation of Einstein coefficients requires, however, the knowledge of transition moments. Hence, in a second step, from the vibration-dependent constants provided by PGopher, we derived the Dunham constants and used them in the RKR\footnote{RKR (v2.0) and LEVEL (v8.0)  are programs developed by Robert J. Leroy, University of Waterloo Chemical Physics Research Reports CP-663 (2007) and CP-657R (2004), http://leroy.uwaterloo.ca/programs/} code to calculate the potential $V_J(r)$. These potentials were injected into the LEVEL\footnotemark[3] program to compute the transition moment matrix elements, $\langle v'J'|Re(r)|v''J''\rangle$, where $Re(r)$ is the transition moment as a function of the internuclear distance $r$, and $|v'J'\rangle$ and $|v''J''\rangle$ are, respectively, the upper and the lower state vibrational wavefunctions. The matrix elements of the R(0) levels were then used as input to PGopher. However, in contrast to \citet{Li2012}, we attempt to consider the rotational dependence of the transition moment matrix elements because we observed transitions involving highly excited rotation levels in order to consider the rotational effect on oscillator strengths. For example, the Einstein coefficient is decreased by $\sim$40\% between the lowest and the highest observed rotational A-X transition. We simultaneously attempt to correct for the Herman-Wallis effect.
In order to approximate the rotational dependence of the transition moment and the Herman-Wallis correction, we approximated the moment matrix elements provided by LEVEL by fitting a polynomial function such that    
\begin{eqnarray}\label{eqn:transpolfit}
 |\langle v'J'|Re(r)|v''J''\rangle | & = & \sum\limits_{i=1}^n  a_i(v',v'',J') J''^i,
\end{eqnarray}
  where the polynomial order $n$ is determined by requiring the $rms$ of the fit to be less than 1\%. Because the Einstein coefficient is proportional to the square of the transition matrix elements, we then correct the Einstein coefficients provided by the PGopher program by the square of the polynomial value for each transition.  We note that it is not obvious that this polynomial fit of singlet-singlet transitions as delivered by LEVEL can be applied to doublet transitions. To test our approach, we therefore synthesize the solar spectrum with and without the polynomial correction. As demonstrated by Fig.~\ref{fig:rot_depend}, the synthetic spectrum of the Sun is improved greatly on the line intensity when corrections are applied.

In a third step, once a first CH linelist had been generated with PGopher, we generated a number of synthetic stellar spectra using Turbospectrum\footnote{Turbospectrum, a LTE radiative transfer code for spectral synthesis in cool stellar atmospheres \citep[see ][]{Alvarez1998, 2012ascl.soft05004P}}, and appropriate MARCS stellar model atmospheres  \citep{Gustafsson2008}. In addition to CH, we included other atomic lines from VALD \citep{Kupka1999,Hill2002,MasseronPhD}, as well as molecular linelists of CN and C$\rm_2$ (B. Plez, private communication) and MgH from the Kurucz linelists.  These spectra were compared with carefully selected observed stellar spectra (see Section \ref{sec:stellar_obs}) in an attempt to identify CH lines as yet unobserved in the laboratory. Thanks to the specific thermodynamic conditions present in cool carbon-star atmospheres, new transitions could indeed be observed and assigned. Although the resolution of spectrographs mounted at telescopes (typically $\lambda/\Delta\lambda=60\,000$, or $\approx 0.4\,{\rm cm}^{-1}$ in our spectral range) does not reach that of laboratory spectrometers ($\approx 10^{-2}\,{\rm cm}^{-1}$), the inclusion of some high rotational transitions only observed  in stellar spectra allows us to improve the determination of the molecular constants in PGopher, with the appropriate weight for each line set by the measurement accuracy. We iterated the process of identification in stellar spectra, and recalculated the list until no new lines could be identified.

Finally, we computed the radiative lifetime of each level ($\tau_{v'J'}$) according to the formula 
\begin{eqnarray}\label{eq:gammarad}
\tau_{v'J'}&=& (\sum_{v''J''}A_{v'J'v''J''})^{-1},
\end{eqnarray}
where $A_{v'J'v''J''}$ is the Einstein coefficient of one transition.
As CH is a molecule known to show predissociation \citep{Herzberg1969}, we also included the predissociation lifetimes, in order to allow for a better estimate of the line broadening parameter $\Gamma_{rad}=1/\tau_{v'J'}+1/\tau_{\rm predissoc}$.  

\section{Observed transitions}\label{sec:obs_trans}
\subsection{$^{12}$CH laboratory data}
Laboratory data was collected from the literature. For the X$^2\Pi$ fundamental state, we used the observations of \citet{Colin2010} of the ro-vibrational transitions in the ACE \citep[Atmospheric Chemical Experiment;][]{Bernath2005} solar spectrum. The quality and resolution of this spectrum is close to laboratory standard ($0.02\,{\rm cm}^{-1}$ in line position accuracy) and the 777 observed transitions allowed an accurate determination of the X$^2\Pi$ state constants up to $v=5$.
For the A$^2\Delta$-X$^2\Pi$ transition, we merged the laboratory measurements of \citet{Bernath1991} and \citet{Zachwieja1995}, with an average accuracy of $0.005\,{\rm cm}^{-1}$ over 665 and 432 observed transitions, respectively.    
The experimental data for the B$^2\Sigma^-$-X$^2\Pi$ transition was taken from \citet{Kepa1996}, \citet{Kumar1998}, and \citet{Bernath1991}, with respective accuracies (and number of lines) of $0.015\,{\rm cm}^{-1}$ (169), $0.01\,{\rm cm}^{-1}$ (73), and $0.005\,{\rm cm}^{-1}$(126). When the same line was present in different studies, we favoured the value from the most accurate measurement. 
For the C$^2\Sigma^+$-X$^2\Pi$ transition, we used the experimental data from \citet{Bembenek1997} (56 lines), \citet{Ubachs1986} (35 lines), and \citet{Heimer1932} (159 lines) for the 0-0 transition with respective accuracies of $0.02\,{\rm cm}^{-1}$, $0.003\,{\rm cm}^{-1}$, $0.2\,{\rm cm}^{-1}$, and \citet{Li1999} (124 lines) for the 1-1 transition with an accuracy of $0.05\,{\rm cm}^{-1}$, and \citet{Herzberg1969} (42 lines) for the 2-2 transition, with an accuracy of $0.2\,{\rm cm}^{-1}$.
Some line assignments from these papers did not correspond to the identification we could make with PGopher. They were changed to the closest matching wavenumber when using the published constant of each author (see Table~\ref{tab:newassignments}).

\begin{table*}
\caption{Modified line assignments for the $\rm ^{12}CH$ transitions. NA stands for lines used in previous studies that we no longer consider in the constant determination process.}
\begin{tabular}{|ll|ll|ll|}
\hline
\multicolumn{2}{c|}{\citet{Colin2010} X-X} & \multicolumn{2}{c}{\citet{Bernath1991} B-X}  & \multicolumn{2}{|c}{\citet{Kepa1996} B-X}\\
\hline
previous   & new                      & previous   & new                               &   previous   & new                      \\
$v'$-$v''$ F($J''$)& $v'$-$v''$ F($J''$)  &      $v'$-$v''$ F($J''$)& $v'$-$v''$ F($J''$)      &     $v'$-$v''$ F($J''$)& $v'$-$v''$ F($J''$)    \\
\hline
1-0 P2ee(1.5)& 1-0 P12ee(1.5) & 1-1 Q12fe(0.5) & 1-1 Q1fe(0.5)                          & 0-1 P2ee(0.5) & 0-1 Q1fe(0.5) \\
1-0 P2ff(1.5)& 1-0 P12ff(1.5) & 1-1 Q2ef(0.5) & 1-1 Q21ef(0.5)                         & 0-1 Q2ef(0.5) & 0-1 Q21ef(0.5)\\
2-1 P2ee(1.5)& 1-0 P12ee(1.5) & 1-1 R1ee(0.5) & 1-1 R21ee(0.5)                         & 1-1 P2ee(0.5) & 1-1 Q1fe(0.5)\\
2-1 P2ff(1.5)& 1-0 P12ff(1.5) &               &                                        & 1-1 Q2ef(0.5) & 1-1 Q21ef(0.5)   \\
2-1 Q2ef(0.5)& 1-0 Q1fe (0.5) &               &                                        &   &  \\
2-1 Q2fe(0.5)& 1-0 Q1ef (0.5) &               &                                        &   &  \\
2-1 R2ee(0.5)& 1-0 R21ee(0.5) &               &                                        &   &  \\
2-1 R2ff(0.5)& 1-0 R21ff(0.5) &               &                                        &   &  \\
4-3 P2ee(1.5)& 1-0 P12ee(1.5) &               &                                        &   &  \\
4-3 R2ff(0.5)& 1-0 R21ff(0.5) &               &                                        &   &  \\
\hline
\multicolumn{2}{c|}{\citet{Zachwieja1995} A-X} & \multicolumn{2}{c}{\citet{Ubachs1986} C-X}  & \multicolumn{2}{|c}{\citet{Bembenek1997} C-X }\\
\hline
previous   & new                      & previous   & new                               &   previous   & new                      \\
$v'$-$v''$ F($J''$)& $v'$-$v''$ F($J''$)  &      $v'$-$v''$ F($J''$)& $v'$-$v''$ F($J''$)      &     $v'$-$v''$ F($J''$)& $v'$-$v''$ F($J''$)    \\
\hline
1-1 R2ff(0.5) & 1-1 R21ff(0.5)        & 0-0 Q1ef(0.5)    & 0-0 Q12ef(0.5)             & 0-0 R21ff(0.5)   &  NA \\
1-1 R2ee(0.5) & 1-1 R21ee(0.5)        & 0-0 Q21fe(0.5)   & 0-0 Q2fe(0.5)              &    &   \\
2-2 R2ff(0.5) & 2-2 R21ff(0.5)        & 0-0 R1ee(0.5)    & 0-0 R12ee(0.5)             &    &   \\
2-2 R2ee(0.5) & 2-2 R21ee(0.5)        &                  &                            &    &   \\
3-3 R2ff(0.5) & 3-3 R21ff(0.5)        &                  &                            &    &   \\
3-3 R2ee(0.5) & 3-3 R21ee(0.5)        &                  &                            &    &   \\
0-1 R2ff(0.5) & 0-1 R21ff(0.5)        &                  &                            &    &    \\
1-2 R2ee(0.5) & 1-2 R21ee(0.5)        &                  &                            &    &    \\
1-2 P12ee(2.5) & NA                  &                   &                            &    &    \\
1-2 P12ff(2.5) & NA                  &                   &                            &    &    \\
\hline                                                                                            
\end{tabular}\label{tab:newassignments}                                                                            
\end{table*}

\subsection{$\rm ^{13}$CH}\label{sec:C13H_labs}
We could not find any observations of the $\rm ^{13}$CH ro-vibrational bands in the literature.
For the A-X, B-X, and C-X transitions, we use the observations of \citet{Zachwieja1997} (601 lines),  \citet{Para1996} (177 lines), and \citet{Bembenek1997} (52 lines), with the respective quoted accuracies of $0.005\,{\rm cm}^{-1}$, $0.015\,{\rm cm}^{-1}$, $0.02\,{\rm cm}^{-1}$. 
We reassigned some lines for $\rm ^{12}$CH in order to match the PGopher identifications (see Table~\ref{tab:13CHreassignments}).

\begin{table*}
\caption{Modified line assignments for the $^{13}$CH transitions.}\label{tab:13CHreassignments}
\begin{tabular}{|ll|ll|ll|}
\hline
\multicolumn{2}{|c}{\citet{Zachwieja1997} A-X} & \multicolumn{2}{|c|}{\citet{Para1996} B-X} & \multicolumn{2}{c|}{\citet{Bembenek1997} C-X} \\ 
\hline
previous             & new               &  previous        & new               &    previous       & new              \\
$v'$-$v''$ F($J''$)    & $v'$-$v''$ F($J''$) & $v'$-$v''$ F($J''$)& $v'$-$v''$ F($J''$) & $v'$-$v''$ F($J''$) & $v'$-$v''$ F($J''$) \\
\hline
1-1 R2ff(0.5)& 1-1 R21ee(0.5) & 0-1 Q12fe(0.5) & 0-1 Q11fe(0.5) & 0-0 R1ee(0.5)  & 0-0 R12ee(0.5)\\
1-1 R2ee(0.5)& 1-1 R21ff(0.5) & 0-1 R2ee(0.5)  & 0-1 R21ee(0.5) &  & \\
2-2 R2ff(0.5)& 2-2 R21ee(0.5) & 1-1 R2ee(0.5)  & 1-1 R21ee(0.5) &  & \\
2-2 R2ee(0.5)& 2-2 R21ff(0.5) & 1-1 Q12fe(0.5) & 1-1 Q11fe(0.5) &  &  \\
3-3 R2ff(0.5)& 3-3 R21ee(0.5) & 1-1 Q2ef(0.5)  & 1-1 Q21ef(0.5) &  &  \\
3-3 R2ee(0.5)& 3-3 R21ff(0.5) &  &  &  & \\
0-1 R2ff(0.5)& 0-1 R21ee(0.5) &  &  &  & \\
0-1 R2ee(0.5)& 0-1 R21ff(0.5) &  &  &  & \\
1-2 R2ff(0.5)& 1-2 R21ee(0.5) &  &  &  & \\
1-2 R2ee(0.5)& 1-2 R21ff(0.5) &  &  &  & \\
2-3 R2ff(0.5)& 2-3 R21ee(0.5) &  &  &  & \\
2-3 R2ee(0.5)& 2-3 R21ff(0.5) &  &  &  & \\
\hline
\end{tabular}
\end{table*}

\subsection{Stellar observations and modelling}\label{sec:stellar_obs}
One may wonder why predissociation lines of CH, a molecule observed in
basically all cool stars, were never detected before.
They are definitely present (see Fig.~\ref{fig:BX}), although never recognized, in stars
as accurately known as the Sun, but such shallow flux depressions, in the
middle of a forest of other atomic and molecular lines, were
never previously identified as distinct spectral features.

Metal-poor stars, on the contrary, display much weaker atomic and
molecular features, allowing a better estimate of the continuum level,
and of possible deviations from it. It is therefore in the spectra of
metal-poor stars, especially those with carbon-enrichment, that the
weak and wide lines caused by CH predissociation are best seen.

\begin{figure*}[h!]
\includegraphics[width=14cm,angle=-90]{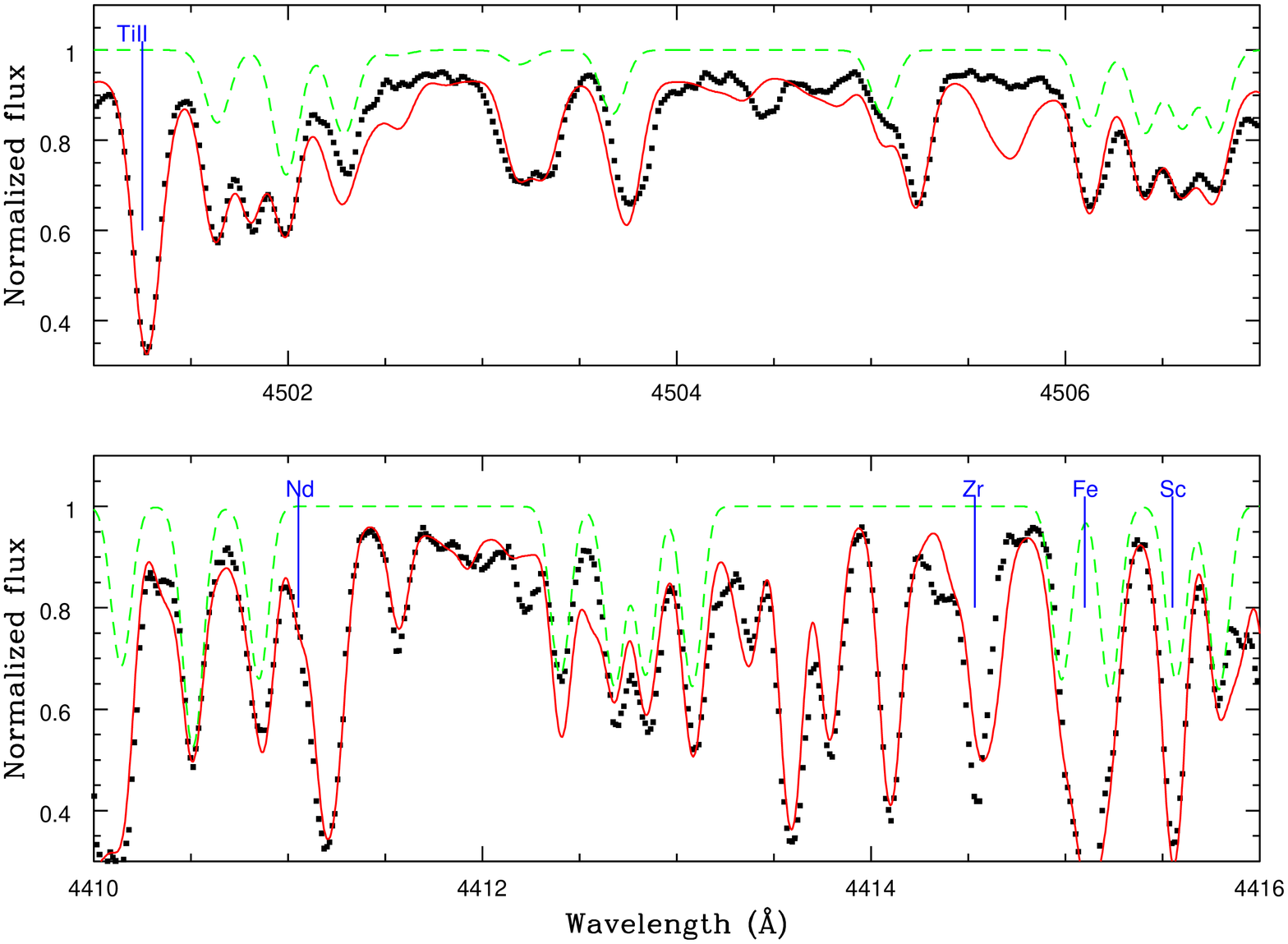}
\caption{Example of stellar spectrum syntheses in the CH A-X transition region. The stellar spectrum is plotted with black points, the red continuous line is the synthesis including both atomic and molecular lines, and the green dashed line is the synthesis with only CH A-X 5-5 transitions in HE~2201-0345  (upper panel) and  CH A-X 4-4 transitions  in CS~22942-019 (lower panel) identified for the first time in a spectrum. We note that some of the mismatches between the synthesis and the observation (e.g. 4505.7\AA, 4412~\AA) can be attributed to inaccuracies of other atomic or molecular linelists.}
\label{fig:AX}
\end{figure*}

Therefore, we use four stellar spectra: CS~22942-019 and HD~224959 both from VLT/UVES ranging from  3300\,\AA\ to 5500\,\AA,  HD~196944 also from VLT/UVES with a spectral coverage of 3050-3850\,\AA\ better suited for the analysis of the C-X band, and a Keck/HIRES spectrum of HE~2201-0345 from 3650\,\AA\ to 4950\,\AA. The resolving power of these instruments, $R=\lambda/\Delta \lambda \approx 60\,000$, combined with the turbulent and thermal Doppler broadening of the lines, leads to an accuracy of typically $0.1\,$\AA\ in line positions, or about  $0.5\,{\rm cm}^{-1}$ in these spectral ranges.  The positions of the predissociation lines of the B-X system are difficult to measure accurately as they are very broad and sometimes even diffuse. We estimate a $10\,{\rm cm}^{-1}$ uncertainty for them (see Sect.~\ref{sec:Bstate}).
The lines were weighted accordingly in PGopher fits. 
The four stars are cool metal-poor C-rich giants, with $T_{\rm eff}\sim 5000\,{\rm K}$. Their spectra show strong CH features, but they are only weakly blended by atomic lines. While CS~22942-019 and HE~2201-0345 have $\rm ^{12}C/^{13}C$ ratios around 10, HD~196944 and HD~224959 have a carbon isotopic ratio of 4 to 5, allowing an easier identification of $\rm^{13}$CH lines. 
The blending of CH features by other atomic and molecular features may
lead to confusion. Therefore we strive to also reproduce the line intensities in order to resolve ambiguities (see Fig.~\ref{fig:AX}). Our MARCS \citep{Gustafsson2008} model atmospheres were specifically tailored to take into account the specific chemical composition (notably C and O), because this affects the thermodynamical structure \citep[see the discussion in][]{MasseronPhD}. The elemental abundances are from \citet{Masseron2010} (except for HE~2201-0345, Masseron et al., in prep.). They are listed in Table~\ref{tab:abund}. 
We note that $\rm ^{13}$CH lines are always weaker than the corresponding $\rm ^{12}$CH lines, because $\rm ^{13}$C is always less abundant than $\rm ^{12}$C. 
Line identifications could only be made for the strongest $\rm ^{13}$CH lines, limiting the number of molecular constants we could derive.

\begin{table*}
\caption{Adopted stellar parameters and abundances for the sample stars.}
\begin{tabular}{|l|c|c|c|c|}
\hline
                         & HD~196944   & HD~224959   &  CS~22942-019  & HE~2201-0345 \\
\hline
$T_{\rm eff}$            & 5250        &   4900      &     5100       & 4800         \\
$\log g$             & 1.7         &   2.0       &     2.5        & 1.9          \\
$\rm [Fe/H]$             & -2.25       &  -2.10       &    -2.50        & -2.10        \\
$\rm [\alpha/Fe]$              & +0.40       &  +0.40     &    +0.40        & +0.40         \\
$\log\rm A(C)$            & 7.35        &  8.15       &    8.10         & 7.95             \\
$\rm ^{12}C/^{13}C$      & 5           &  4          &    12          & 10              \\
$\log\rm A(N)$            & 7.05        &  7.60        &   6.40          & 6.70             \\
$\log\rm A(O)$            & 6.8       &  7.7        &   6.7          & 7.5             \\
$\rm [Sr,Y,Zr,Ba,La,Pb/Fe]$ & 1.5         &  2.0        &   1.5         &  1.3            \\
\hline  
 \end{tabular}\label{tab:abund}                                                                                  
\end{table*}

The positions of newly identified lines of the $^{12}$CH A-X, B-X, and C-X transitions in our stellar spectra are reported in Tables~\ref{tab:AX_lines}, \ref{tab:BX_lines}, and \ref{tab:CX_lines}, while those for $^{13}$CH are reported in Tables~\ref{tab:AX_C13H_lines}, \ref{tab:BX_C13H_lines}, and \ref{tab:CX_C13H_lines} (available online). \\
For the A-X transition, we observed levels up to $J'' = 36.5$ for $v' = 0$, $J'' = 32.5$ for $v' = 1$, $J'' = 31.5$ for $v' = 2$, and  $J'' = 27.5$ for $v' = 3$, very close to the dissociation limit, while only levels up to $J'' = 27.5$ for $v' = 0$, $J'' = 19.5$ for $v' = 1$, $J'' = 12.5$ for $v' = 2$, and $J'' = 10.5$ for $v' = 3$ were measured by \citet{Zachwieja1995}. We were able to determine for the first time the constants for the A$^2\Delta$ $v=4$ and $v=5$ states. 
For the B-X transition (Table~\ref{tab:BX_lines}), we observed transitions from higher $J''$ levels than \citet{Kepa1996} and \citet{Bernath1991}  ($J'' < 15.5$), and than \citet{Kumar1998} ($J'' < 21.5$). 
For the C-X transition, we also observe lines to higher rotational levels in our CEMP star spectrum than in laboratory spectra. There is an excellent agreement with the predicted line positions derived from the laboratory experiments.

%%%%%%%%%%%%%%%%%%%%%%%%%%%%%%%%%%%%%%%%%%%%%%%%%%%%%%%%%%%%%%%%%%%%%%%%%%%%%%%%%%%%
\section{Results}\label{sec:results}
\subsection{Revised molecular constants for $^{12}$CH}
After merging all data of all transitions (i.e. ro-vibrational, A-X, B-X, and C-X) from laboratory and from stellar spectra observations into PGopher, weighted according to their respective accuracy, we were able to derive a single set of new molecular constants for the four lowest doublet electronic states of CH. 
The constants we provide in this section are consistent with the definition of the Hamiltonian of \citet{Brown1979}: $B_v$ is the rotational constant; $D_v$, $H_v$, $L_v$, and $M_v$ are the centrifugal distortion constants. $A_v$ is the spin-orbit constant; $\gamma_v$ and $\gamma_{Dv}$ are the spin-rotation constant and its centrifugal distortion; and $p_v$ and $q_v$ are the $\Lambda$-doubling constants.
For the computation of Einstein coefficients and level radiative lifetime, we used the electronic-transition moments of \citet{vanDishoeck1987}. For the ro-vibrational transitions, we used the dipole moment function calculated by \citet{Hettema1994}. We scaled the electronic-transition moments to match  the experimental lifetime measurements: $\rm \tau_{00} = 535$~ns for the A$^2\Delta v=0$ level \citep{Luque1996_AX},  $\tau_{00} = 325$~ns for the B$^2\Sigma^- ~v=0$ level \citep{Luque1996_BX}, and $\tau_{00} = 110$~ns for the C$^2\Sigma^+$ $v=0$ level \citep{Brzozowski1976}. 
% The polynomial coefficients used in PGopher to describe the dependence of the transition moments with $J$ (see Eq.~\ref{eqn:transpolfit}) are given in Table~\ref{tab:transpolcoef}.   We provide the same Einstein coefficients, and hence $gf$-values, for both $\rm ^{12}CH$ and $\rm ^{13}CH$.

\subsubsection{The X$^2\Pi$ fundamental state }
Thanks to the supplementary data offered by our spectra in the optical region, we could improve the ground state constants. 
The constants we obtained are presented in Table~\ref{tab:X_const}. They are consistent with the work of \citet{Colin2010}, as we included their data in our fit. Moreover, thanks to our observations of high J A-X transitions (see Fig.~\ref{fig:AX}), we could improve the accuracy of the high-order constants, notably of $L_v$.  

\begin{table*}
\caption{Constants (in cm$^{-1}$) for the X$^2\Pi$ state of $^{12}$CH. Uncertainties in parentheses are one standard deviation in units of the last quoted digit. The second line contains the values of  \citet{Colin2010}. }\label{tab:X_const}
\begin{tabular}{lllllll}
\hline
Constant               & $v=0$         & $v=1$               & $v=2$           & $v=3$              & $v=4$                &$v=5$                       \\
\hline                                           
$T_v$            & 0.0              & 2732.97900(130)  & 5339.9051(20)    & 7822.2222(28)    & 10181.0060(41)   & 12416.755(24) \\ 
~                & 0.0              & 2732.97813(94)   & 5339.9036(15)    & 7822.2184(18)    & 10180.9996(22)   & 12416.786(10)   \\ 
$B_v$            & 14.192249(57)    & 13.6617020(560)  & 13.135783(58)    & 12.613439(67)    & 12.092930(100)   & 11.57387(52)    \\ 
~                & 14.192393(61)    & 13.6617896(53)   & 13.135809(50)    & 12.613441(50)    & 12.092993(54)    & 11.57300(23)    \\ 
$A_v$            & 28.14675$^a$     & 28.33833$^a$     & 28.5312(78)       & 28.652(14)       & 28.909(30)       & 28.64(49)   \\ 
~                & 28.14675$^a$     & 28.33833$^a$     & 28.6160(510)       & 28.765(48)       & 28.968(45)       & 29.17$^b$       \\
$\gamma_v$$\times$10$^2$      & -2.624(33)       & -2.424(32)       & -2.267(37)       & -2.136(46)       & -1.954(62)       & -1.86(16)      \\ 
          & -2.853(86)       & -2.654(85)       & -2.462(83)       & -2.323(81)       & -2.138(80)       & -1.933(70)      \\ 
$p_v$$\times$10$^2$           & 3.358(29)        & 3.174(24)        & 3.085(34)        & 2.987(48)        & 2.779(95)       & 2.580(150)      \\ 
         & 3.464(36)        & 3.260(34)        & 3.119(36)        & 2.967(42)        & 2.836(52)        & 2.673(62)       \\ 
$q_v$$\times$10$^2$           & 3.8684(20)       & 3.7290(17)       & 3.5902(23)       & 3.4487(35)       & 3.3104(69)       & 3.1724(75)       \\ 
         & 3.8660(31)       & 3.7285(29)       & 3.5919(32)       & 3.4541(36)       & 3.3179(43)       & 3.1776(32)      \\ 
$D_v$$\times$10$^3$          & 1.46079(32)      & 1.43776(32)      & 1.41641(32)      & 1.39722(35)      & 1.37980(56)      & 1.3740(25)       \\ 
          & 1.46112(37)      & 1.43742(31)      & 1.41553(31)      & 1.39612(32)      & 1.37911(35)      & 1.3692(11)      \\ 
$H_v$$\times$10$^7$            & 1.1381(71)       & 1.1061(68)       & 1.0717(64)       & 1.0295(64)        & 0.953(10)       & 0.90 $^b$      \\ 
          & 1.1064(75)       & 1.0625(85)       & 1.0212(97)       & 0.9790(100)        & 0.917(10)        & 0.88$^b$        \\ 
$L_v$$\times$10$^{12}$            & -12.00(54)       & -12.12(47)       & -12.65(41)       & -13.52(37)       & -13.48(68)       & -13.5$^b$             \\ 
       & -3.40(177)         & -4.34(162)      & -5.69(147)       & -7.55(129)       & -9.27(111)       & -11.6$^b$       \\ 
$p_{Dv}$$\times$10$^5$         & -0.853(79)       & -0.728(68)       & -0.685(71)       & -0.532(91)       & -0.52(17)       & -0.6$^b$             \\ 
          & -1.010(370)      & -0.840(340)      & -0.730(320)      & -0.630(290)      & -0.60(27)      & -0.55$^b$       \\ 
$q_{Dv}$$\times$10$^5$         & -1.524(10)      & -1.4865(94)      & -1.4453(94)      & -1.394(12)      & -1.344(26)      & -1.35$^b$       \\ 
          & -1.590(20)     & -1.5500(220)     & -1.5090(240)     & -1.461(24 )     & -1.4090(25)     & -1.35$^b$       \\ 
$q_{Hv}$$\times$10$^9$          & 3.08(13)        & 2.94(11)        & 2.790(10)        & 2.60(11)        & 2.30(24)        & 2.0$^b$            \\ 
         & 4.06(57)       & 3.79(53)       & 3.560(480)       & 3.30(42)       & 2.90(37)       & 2.52$^b$        \\ 
$\gamma_{Dv}$$\times$10$^5$    & 0.99(15)         & 0.90(14)         & 0.88(13)         & 0.91(13)        & 0.81(14)        & 0.8$^b$             \\ 
          & 1.78(24)         & 1.64(22)         & 1.53(21)         & 1.50(19)         & 1.33(18)         & 1.29$^b$        \\ 
\hline                                                             
\end{tabular}
\tablefoot{$^a$ fixed to \citet{Jackson2008}\\
           $^b$ kept fixed}
\end{table*}

\subsubsection{The A$^2\Delta$ state}
Table~\ref{tab:A_const} contains the constants we derive for the A$^2\Delta$ state. We were able to determine for the first time the constants for the A$^2\Delta$ $v=4$ and $5$ levels (Fig.~\ref{fig:AX}). For the lower vibrational levels, our constants agree well with the ones of \citet{Zachwieja1995} because we used of their laboratory measurements. The small differences come from the fact that we obtained different constants for the ground state. However, in contrast to \citet{Zachwieja1995}, we could not recover the $\Lambda$-doubling constants $p_v$ and $q_v$ similar to \citet{Bernath1991}. The resulting $rms$ difference of the observed minus calculated line positions is $1.46\times 10^{-2}\,{\rm cm}^{-1}$, which is very satisfactory for studies of stellar spectra. 
\begin{table*}
\caption{Constants (in cm$^{-1}$) for the A$^2\Delta$ state. Uncertainties in parentheses are one standard deviation in units of the last quoted digit. }\label{tab:A_const}
\begin{tabular}{lllllll}
\hline
Constant         & $v=0$                 & $v=1$                & $v=2$                & $v=3$               & $v=4$               & $v=5$     \\
\hline
$T_v$            & 23173.45580(110)     & 25913.7404(21)     & 28460.7224(32)     & 30794.1049(45)     & 32885.355(84)  & 34695.691(73) \\ 
~                & 23173.45853(56)$^a$ & 25913.7454(11)$^a$ & 28460.7406(24)$^a$ & 30794.1115(20)$^a$ & ~                 & ~               \\ 
$B_v$            & 14.579064(58)       & 13.911205(75)      & 13.18806(11)     & 12.37277(17)       & 11.4667(21)       & 10.33371(21)  \\ 
~                & 14.579083(47)$^a$   & 13.910878(60)$^a$  & 13.18663(12)$^a$   & 12.38208(11)$^a$   & ~                 & ~               \\ 
$D_v$$\times$10$^3$     & 1.56621(34)         & 1.60903(46)        & 1.6909(11)         & 1.8275(21)         & 2.171(15)         & 2.455(19)       \\ 
~                & 1.56604(30)$^a$     & 1.60408(88)$^a$    & 1.6689(12)$^a$     & 1.8157(12)$^a$     & ~                 & ~               \\ 
$H_v$$\times$10$^8$     & 9.962(61)          & 8.94(11)           & 8.60(27)           & 6.30(62)         &   32.5(38)        & -38.9(53)       \\ 
~                & 9.761(72)$^a$       & 6.98(46) $^a$      & ~                  & ~                  & ~                 & ~               \\ 
$L_v$$\times$10$^{11}$  & -2.817(33)           & -5.274(69)          &-11.88(17)           & -20.634(50)        &    -93.2(32)     & -127.2(30)               \\ 
~                & 2.666(55)$^a$       &  3.09(77)$^a$      & ~                  & ~                  & ~                 & ~               \\ 
$A_v$            & -1.10150(140)       &-1.0721(20)         &-1.0402(26)         & -0.9934(38)        &-0.95$^b$           & -0.9$^b$    \\
                 & -1.10088(88)$^a$    &-1.0697(13)$^a$     & -1.0392(27)$^a$    & -0.9982(27)$^a$    & ~                 & ~               \\ 
$\gamma_v$$\times$10$^2$ & 4.171(37)          & 3.979(49)          & 3.821(53)          & 3.370(75)          & 3.21(27)          & 3.0$^b$          \\ 
~                & 4.217(25)$^a$       & 4.000(20)$^a$      & 3.778(44)$^a$      & 3.518(37)$^a$      & ~                 & ~               \\ 
$\gamma_{Dv}$$\times$10$^5$& -0.730(150)      & -0.790(270)        & ~                  & ~                  & ~                 & ~               \\ 
~                   & -0.925(65)$^a$   & -0.871(92)$^a$     & ~                  & ~                  & ~                 & ~               \\ 
$p_v$$\times$10$^7$     & ~                   & ~                  & ~                  & ~                  & ~                 & ~               \\ 
                 & 6.6(14)$^a$         & 13.0(74)$^a$       & 28(11)$^a$         & 58(88)$^a$         & ~                 & ~               \\ 
$q_v$$\times$10$^8$     & 4.53(88)            & 14.1(53)           & ~                  & ~                  & ~                 & ~               \\ 
~                & -2.89(36)$^a$       & -17.5(24)$^a$      & -35.(52)$^a$       & -133.(60)$^a$      & ~                 & ~               \\ 
\hline
\end{tabular}
\tablefoot{ $^a$ Constants by \citet{Zachwieja1995}. $^b$ kept fixed}
\end{table*}
Because some of the A levels are expected to predissociate (see Sect.~\ref{sec:methodo}), their lifetime should be shorter than  their radiative lifetime \citep{Brzozowski1976}. However, we did not observe any increased broadening for the A-X lines in our stellar spectra. The predissociation broadening is dominated by thermal and turbulence broadening in our target stars. Nevertheless, the predissociation lifetimes are included in our linelist.

\subsubsection{The B$^2\Sigma^-$ state}\label{sec:Bstate}
Table~\ref{tab:B_const} presents the constants we have derived for the B$^2\Sigma^-$ state. The constants are in good agreement with \citet{Kepa1996}. 
\begin{table*}
\caption{Constants (in cm$^{-1}$) for the B$^2\Sigma^-$ state. Uncertainties in parentheses are one standard deviation in units of the last quoted digit.}\label{tab:B_const}
\begin{tabular}{llllll}
\hline
Constant     & $v=0$ & $v=1$  \\
\hline
$T_v$            & 25712.5074(24)     & 27507.5216(24)  \\ 
~                & 25712.5113(12)$^a$ & 27507.5250(8)$^a$            \\ 
$B_v$            & 12.640320(220)     & 11.15815(48)    \\ 
~                & 12.640159(66)$^a$  & 11.15788(14)$^a$            \\ 
$D_v$$\times$10$^3$     & 2.13110(560)       & 3.1680(260)       \\ 
~                & 2.12954(94)$^a$    & 3.1260(60)$^a$            \\ 
$H_v$$\times$10$^7$     & -4.930(540)        & -14.90(500)       \\ 
~                & -4.114(36)$^a$     & -30.22(74)$^a$            \\ 
$L_v$$\times$10$^{10}$  & 9.7(21)            & 0.0             \\ 
~                & 0.0$^a$                & 0.0$^a$             \\ 
$M_v$$\times$10$^{12}$  & -2.70(27)          & 0.0               \\ 
~                & 0.0$^a$                & 0.0$^a$             \\ 
$\gamma_v$$\times$10$^2$  & -2.938(47)       & -2.1(46)      \\ 
~                  & -2.702(16)$^a$   & -2.216(13)$^a$            \\ 
$\gamma_{Dv}$$\times$10$^5$  & 2.08(39)       &    0.0$^a$   \\ 
~                  & 0.0$^a$   &         0.0$^a$   \\ 
\hline
\end{tabular}
\tablefoot{$^a$ Constants derived by \citet{Kepa1996}. }
\end{table*}

\begin{table}
\caption{Radiative lifetimes from this work and experimental lifetimes from \citet{Luque1996_BX} of the B$^2\Sigma^-$ levels in ns. $^a$ extrapolated value from astrophysical observations. Predissociation onset is the limit of the lowest level with evidence for predissociation and predissociation barrier indicates the limit of the uppest level with evidencce for predissociation.
\label{Tab:width}}
\begin{tabular}{r|rr|rr|}
\hline
     &\multicolumn{2}{c}{$v=0$} & \multicolumn{2}{c}{$v=1$} \\
\hline
  J  & $\tau_{rad}$  & $\tau_{exp}$&      $\tau_{rad}$ & $\tau_{exp}$    \\  
0.5  & 328  &        &   472 &                                  \\
1.5  & 329  &        &   474 &                                 \\
2.5  & 330  & 328    &   476 &                                 \\
3.5  & 332  & 345    &   480 & 396                             \\
4.5  & 335  & 328    &   486 & 410                             \\
5.5  & 338  & 346    &   493 & 405                             \\
     &      &        &    \multicolumn{2}{c|}{predissociation onset} \\
6.5  & 343  & 356    &   502  & 390                             \\
7.5  & 348  & 354    &    512 &     5.0       \\             
8.5  & 354  & 360    &    526 &     0.350        \\          
9.5  & 361  & 345    &    543 &     0.075       \\           
10.5 & 370  & 352    &    563 &     0.015       \\           
11.5 & 380  & 364    &    589 &     0.005       \\           
12.5 & 392  & 362    &    620 &     0.0013       \\          
13.5 & 406  & 395    &        &     0.0004       \\          
14.5 & 423  & 435    &        &     0.0002$^a$     \\             
     & \multicolumn{2}{c|}{predissociation onset} &  & \\
15.5 & 443  & 140      &       &     0.00001$^a$   \\    
16.5 & 468  & 70       &       &     0.000003$^a$   \\      
17.5 & 498  & 0.6      &       &     1.0E-6$^a$     \\   
18.5 & 535  & 0.045    &       &     3.0E-7$^a$   \\     
19.5 & 582  & 0.010    &       &     1.0E-7$^a$   \\     
20.5 & 642  & 0.003    &       &     3.0E-8$^a$   \\     
21.5 & 723  & 0.0004   &       &                  \\
\multicolumn{5}{c}{-- predissociation barrier --} \\
\hline                                    
\end{tabular}
\end{table}

\begin{figure}[h!]
\includegraphics[width=7cm,angle=-90]{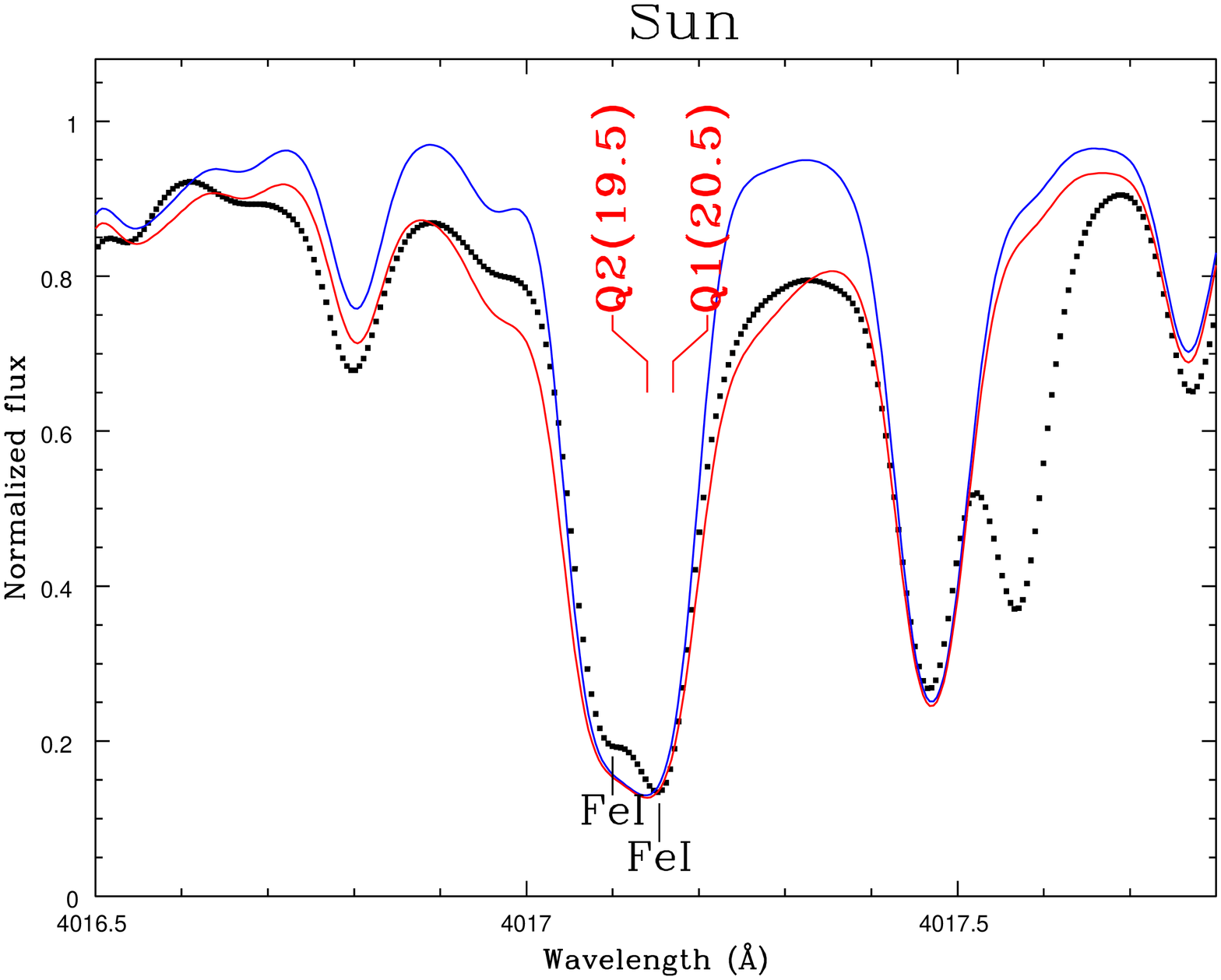}
\includegraphics[width=7cm,angle=-90]{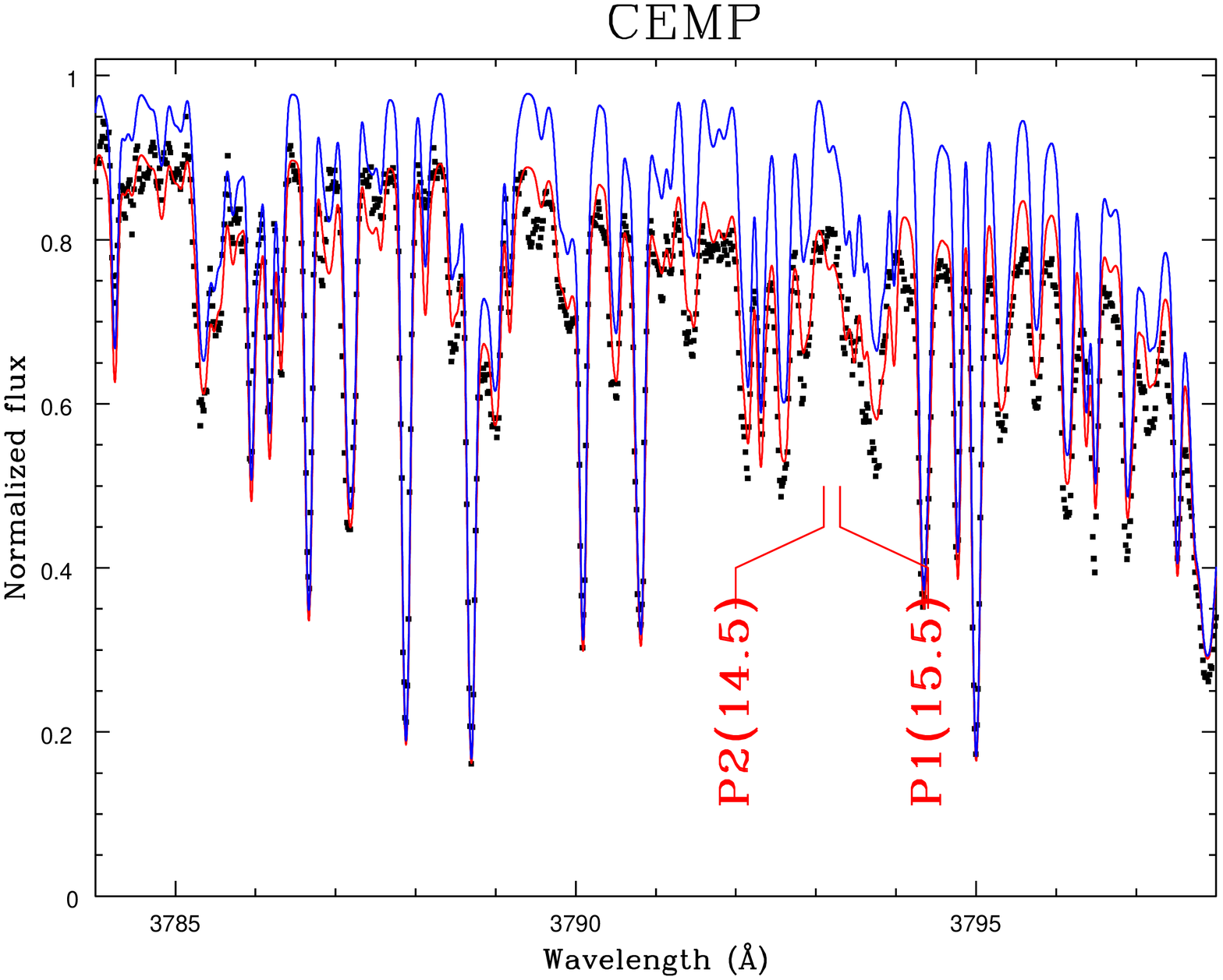}
\caption{Top panel: example of a synthesis of the  solar photospheric spectrum (dotted black) including (red) and omitting (blue) predissociation lines of the B-X system. Bottom panel: Same for the CEMP star CS~22942-019. Predissociation lines, as well as many other CH lines, are present and contribute to the flux depression. In the case of the CEMP star, the presence of several extremely broad CH predissociation lines is so important that it acts as a pseudo-continuum. }
\label{fig:BX}
\end{figure}

There are a number of high $J$ B-X predissociation lines that can be seen even in the solar spectrum (Fig.~\ref{fig:BX}). However, as already noted by \citet{Kumar1998}, their positions are not well defined. These lines do not help to improve the accuracy of the B state constants, but nevertheless we include them in our final linelist.
Predissociative lifetimes of the corresponding levels are listed in Table~\ref{Tab:width}. 
They are in good agreement with the laboratory observations of \citet{Kumar1998}, and \citet{Luque1996_BX}.
Because of the increasing broadening of the lines with rotational quantum number, the identification of the lines becomes more and more uncertain as $J$ increases. Consequently, the identification of some levels is ambiguous and the exact dissociation energy of the B state is very difficult to determine.  
We set the height of the predissociation barrier according to our new line identifications, that include the observed $0-0~R1(21.5)$ line, with an upper energy level of 31474~cm$^{-1}$. The height of the barrier is thus at least $31474 - 29374 = 2100$~cm$^{-1}$, with $D_e = 29374\pm10$~cm$^{-1}$ \citep{Kumar1998}, almost twice as high as the value of 1150 cm$^{-1}$ predicted by \citet{vanDishoeck1987}.

Assuming this barrier height, B$^2\Sigma^-$  $v=1$ levels have been included up to $J''=20.5$ in our list, although we were able to identify B-X transitions only up to $J''=16.5$ because the combination of low gf values and large predissociation broadening limits the identification.   However, we stress that these very broad lines act as a pseudo-continuum and must be taken into account in spectrum synthesis (see Fig.~\ref{fig:BX}).

An atlas of predissociation lines for some typical spectra (the Sun, Arcturus, and several carbon-enhanced metal-poor stars) is presented in the Appendix.

\subsubsection{The C$^2\Sigma^+$ state}
\begin{figure}[h!]
\includegraphics[width=7cm,angle=-90]{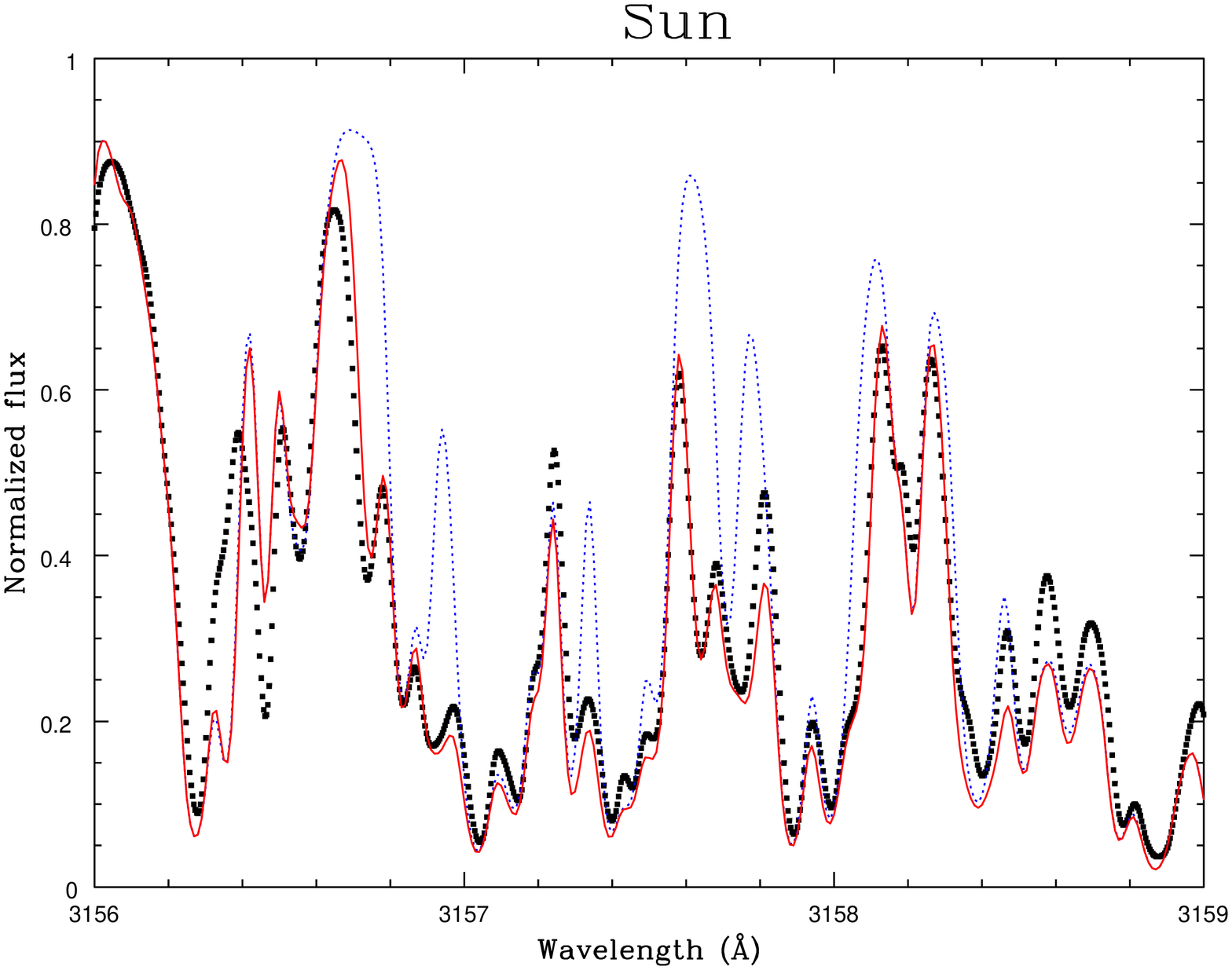}
\caption{Observed solar photospheric spectrum (black dots) and synthetic spectra using the former CX linelist (blue dotted line) from \citet{Jorgensen1996} and our C-X linelist (red solid line).}
\label{fig:CH-CX_Sun}
\end{figure}

We were able to identify C-X lines in our CEMP star spectrum at higher $J$ than in any published laboratory works so far. As shown in Table~\ref{Tab:C_const}, there is an excellent agreement of their measured positions with the predicted values derived from the laboratory constants. As our identifications carry a lower accuracy than laboratory measurements, we did not attempt to use these lines in our global fit. However, thanks to the inclusion of more recent laboratory data from \citet{Li1999} and \citet{Bembenek1997} compared to \citet{Jorgensen1996}, we were able to improve the C-X linelist (see Fig.~\ref{fig:CH-CX_Sun}).\\
\begin{table*}
\caption{Constants (in cm$^{-1}$) for the C$^2\Sigma^+$ state. Uncertainties in parentheses are one standard deviation in units of the last quoted digit. 
}\label{Tab:C_const}
\begin{tabular}{llllll}
\hline
Constant     & v=0  & v=1 & v=2  \\
\hline                          
$T_v$            & 31791.6479(16)   &   34403.310(21)     & 36772.842(54)   \\ 
~                & 31791.6467(8)$^a$ & 34403.106(21)$^a$ & 36772.825(24)$^a$ \\
$B_v$            & 14.255724(90)     & 13.50577(43)      & 12.6046(29)     \\ 
~                & 14.255908(10)$^a$ & 13.51614(66)$^a$  & 12.6076(18)$^a$  \\
$D_v$$\times$10$^3$     & 1.59375(97)       & 1.6336(21)        & 1.918(37)       \\ 
~                & 1.59511(46)$^a$   & 1.7305(55)$^a$    & 2.013(27)$^a$   \\
$H_v$$\times$10$^8$     & 7.45(29)          & -14.7(27)          & -73.(10)        \\ 
~                & 7.67(14)$^a$      & 14.7(16)$^a$      & 0.0$^a$             \\ 
$L_v$$\times$10$^{11}$  & -3.75(26)         & 0.0               & 0.0                 \\
~                & -3.73(12)$^a$     & -28.1(15)$^a$     & 0.0$^a$             \\ 
$\gamma_v$$\times$10$^2$  & 4.126(58)       & 3.320(120)        & 3.80(10)        \\ 
~                  & 4.320(32)$^a$   & 3.650(250)$^a$      & 3.53(43)$^a$    \\
$\gamma_{Dv}$$\times$10$^5$  & -1.66(62)    & 0.0               & 0.0              \\ 
~                & -3.86(51)$^a$     & -1.32(75)$^a$     & 0.0$^a$             \\ 
$\gamma_{Hv}$$\times$10$^8$  & 0.0          & 0.0               & 0.0             \\ 
~                & 3.09(87)$^a$      & 0.0$^a$               & 0.0$^a$             \\
\hline
\end{tabular}
\tablefoot{ $^a$ Values from \citet{Li1999}.}
\end{table*}

The C state is also known to predissociate, but the predissociation lifetimes are larger than for the B state \citep{Brzozowski1976}. We could not detect any increase in the line broadening in our stellar spectra, but we nevertheless include the broadening lifetime in our list using the corresponding predissociation lifetime measurements of \citet{Brzozowski1976} and \citet{Ubachs1986}. 

\subsection{Revised molecular constants for $\rm ^{13}$CH}

Tables ~\ref{tab:C13X_const}, \ref{tab:C13A_const}, \ref{tab:C13B_const}, and \ref{tab:C13C_const} present the molecular constants we derive for the four lowest doublet electronic states of $\rm ^{13}$CH.
The experimental data were taken from \citet{Zachwieja1997},  \citet{Para1996}, and \citet{Bembenek1997} (see Sect.~\ref{sec:C13H_labs}).
Because the abundance of $\rm ^{13}$C is lower than $\rm ^{12}$C in stellar atmospheres, we were not able to observe transitions to much higher $J$ values than in the laboratory, as we did for $\rm ^{12}$CH.  Although the lines observed in stellar spectra contribute to improve the accuracy of high-order $J$ constants, the major improvements in the constants come from our global fit of all the laboratory  data for the three electronic transitions (A-X, B-X, and C-X), which had never been done previously; for example we were able to recover experimentally for the first time the spin-orbit constant ($A_v$) of the fundamental level (Table~\ref{tab:C13X_const}).

We were also able detect for the first time predissociation lines for the B state of $\rm ^{13}CH$. Because the potential and the energy levels are not strictly identical to the $\rm ^{12}CH$ case, predissociation may occur for other levels and with different lifetimes. The quality of our spectra, the presence of blends with other lines, and the diffuse character of the predissociation lines did not allow us to measure any significant difference in the widths of the predissociated lines of $\rm ^{12}CH$ and $\rm ^{13}CH$. Therefore,  the same predissociation lifetimes as for $\rm ^{12}CH$ were adopted.
For the C state, only the constants for the $v=0$ level could be determined as we could not assign any lines linking to other vibrational levels.
Improvements of the  $\rm ^{13}$CH constants could certainly  be obtained with the identification of the infrared ro-vibrational X-X transition.

\begin{table*}
\caption{Constants (in cm$^{-1}$) for the  X$^2\Pi$ state of $\rm ^{13}$CH. Uncertainties in parentheses are one standard deviation in units of the last quoted digit. }\label{tab:C13X_const}
\begin{tabular}{lllll}
\hline
Constant            & $v=0$                & $v=1$          & $v=2$                     & $v=3$           \\
\hline 
$T_v$            & 0.0              & 2725.1667(24)    & 5325.0275(46)    & 7801.0058(58)   \\ 
                 & 0.0              & 2725.1644(12)$^a$& 5325.0249(19)$^a$& 7800.9918(48)$^a$   \\ 
$B_v$            & 14.108424(100)   & 13.582530(110)   & 13.061290(190)   & 12.54273(26)    \\ 
                 & 14.108210(43)$^a$& 13.582435(43)$^a$& 13.061122(62)$^a$& 12.54347(19)$^a$    \\ 
$A_v$            & 28.1426(81)      & 28.3355(87)      & 28.508(21)       & 28.701(13)      \\ 
                 & 28.14643$^{a,b}$ & 28.3384$^{a,b}$  & 28.5243$^{a,b}$  & 28.7041$^{a,b}$         \\ 
$\gamma_v$$\times$10$^2$ & -2.514(59)      & -2.308(53)       & -2.146(66)       & -1.96(13)       \\ 
                  & -2.547(22)$^a$  & -2.363(20)$^a$   & -2.177(27)$^a$   & -2.0728$^{a,b}$        \\ 
$p_v$$\times$10$^2$     & 3.304$^c$        & 3.243(59)        & 2.969(51)        & 2.665(98)       \\ 
                 & 3.313(15)$^a$    & 3.190(13)$^a$    & 3.019(30)$^a$    & 2.843(62)$^a$       \\ 
$q_v$$\times$10$^2$     & 3.823672$^c$     & 3.6916(65)       & 3.5491(85)       & 3.4060(150)       \\ 
                 & 3.8190(12)$^a$   & 3.6849(12)$^a$   & 3.5521(18)$^a$   & 3.4163(40)$^a$      \\ 
$D_v$$\times$10$^3$     & 1.44443(72)      & 1.42004(84)      & 1.40030(270)     & 1.3607(27)      \\ 
                 & 1.44440(29)$^a$  & 1.42186(32)$^a$  & 1.40001(43)$^a$  & 1.3787(17)$^a$      \\ 
$H_v$$\times$10$^7$     & 1.1380(180)      & 1.0240(190)      & 1.08(11)         & 1.0$^b$       \\ 
                 & 1.1457(68)$^a$   & 1.1272(66)$^a$   & 1.0731$^{a,b}$   & 1.0284$^{a,b}$       \\ 
$L_v$$\times$10$^{11}$  & -1.36(14)        & 0.0              & 0.0              & 0.0             \\ 
                 & 1.426(52)$^a$    & 1.386$^{a,b}$    & 1.397$^{a,b}$    & 1.446$^{a,b}$           \\ 
$p_{Dv}$$\times$10$^6$  & -8.51(56)        & -12.6(36)        & 0.0              & 0.0             \\ 
                 & -8.91(40)$^a$    & -7.92$^{a,b}$    & -6.93$^{a,b}$    & -5.94$^{a,b}$           \\ 
$q_{Dv}$$\times$10$^5$  & -1.5100(63)      & -1.604(71)       & -1.373(85)       & 0.0             \\ 
                 & -1.4792(64)$^a$  & -1.4448(52)$^a$  & -1.4101$^{a,b}$  & -1.3717$^{a,b}$         \\ 
$q_{Hv}$$\times$10$^9$  & 3.110(120)       & 7.0(17)          & 0.0              & 0.0             \\ 
                 & 2.724(81) $^a$   & 2.673$^{a,b}$    & 2.580$^{a,b}$    & 2.488$^{a,b}$           \\ 
$\gamma_{Dv}$$\times$10$^6$  & 5.8(22)     & 0.0              & 0.0              & 0.0             \\ 
                      & 8.45(74)$^a$& 8.185$^{a,b}$    & 8.135$^{a,b}$    & 8.354$^{a,b}$           \\ 
\hline
\end{tabular}
\tablefoot{$^a$ Constants derived by \citet{Zachwieja1997}. $^b$ kept fixed. $^c$ taken from \citet{Steimle1986}. }
\end{table*}

\begin{table*}
\caption{Constants (in cm$^{-1}$) for the A$^2\Delta$ state of $\rm ^{13}$CH. Uncertainties in parentheses are one standard deviation in units of the last quoted digit. }\label{tab:C13A_const}
\begin{tabular}{lllll}
\hline
Constant       & $v=0$                & $v=1$              & $v=2$               & $v=3$        \\
\hline
$T_v$            & 23173.84500(200)     & 25906.4648(33)       & 28447.0001(50)       & 30775.3828(78)  \\ 
                 & 23173.84460(69)$^a$  & 25906.4596(15)$^a$   & 28446.9953(30)$^a$   & 30775.3748(59)$^a$  \\ 
$B_v$            & 14.492570(100)       & 13.830140(110)       & 13.11333(140)        & 12.31697(34)    \\ 
                 & 14.492358(41)$^a$    & 13.830193(54)$^a$    & 13.113259(88)$^a$    & 12.31700(24)$^a$    \\ 
$A_v$            & -1.09930(270)        & -1.0741(33)          & -1.0312(38)          & -1.0000(68)     \\ 
                 & -1.09926(90)$^a$     & -1.0735(12)$^a$      & -1.0339(35)$^a$      & -0.9981(38)$^a$     \\ 
$\gamma_v$$\times$10$^2$& 4.211(66)            & 4.056(77)            & 3.717(78)            & 3.530(160)        \\ 
                 & 4.184(24)$^a$        & 3.999(28)$^a$        & 3.783(48)$^a$        & 3.470(53)$^a$       \\ 
$D_v$$\times$10$^3$     & 1.54753(74)          & 1.58069(82)          & 1.64999(87)          & 1.7963(34)      \\ 
                 & 1.54745(28)$^a$      & 1.58442(54)$^a$      & 1.65055(48)$^a$      & 1.7972(22)$^a$      \\ 
$H_v$$\times$10$^8$     & 9.610(180)           & 5.05(18)             & 0.0                  & 0.0             \\ 
                 & 9.664(65)$^a$        & 6.92(20)$^a$         & 0.0$^a$                  & 0.0$^a$             \\ 
$L_v$$\times$10$^{11}$  & -2.61(14)            & 0.0                  & 0.0                  & 0.0             \\ 
                 &  2.667(50)$^a$       & 2.54(24)$^a$         & 0.0$^a$                  & 0.0$^a$             \\ 
$\gamma_{Dv}$$\times$10$^5$  & -1.140(220)     & -1.610(270)          & 0.0                  & 0.0             \\ 
                      & -0.907(75)$^a$  & -0.714(89)$^a$       & 0.0$^a$                  & 0.0$^a$             \\ 
$p_v$$\times$10$^7$     &      0.0             &     0.0              &    0.0               &    0.0          \\ 
                 &   5.7(18)$^a$        &   10.4(74)$^a$       &   20.5$^{a,b}$       &   27.5$^{a,b}$              \\ 
$q_v$$\times$10$^8$     &      0.0             &     0.0              &    0.0               &    0.0          \\ 
                 &   -2.84(41)$^a$      &  -16.3(23)$^a$       &   -32.4$^{a,b}$      &    -47.3$^{a,b}$             \\ 
\hline
\end{tabular}
\tablefoot{$^a$ Constants derived by \citet{Zachwieja1997}. $^b$ kept fixed in \citet{Zachwieja1997}.}
\end{table*}

\begin{table*}
\caption{Constants (in cm$^{-1}$) for the  B$^2\Sigma^-$ state of $\rm ^{13}$CH. Uncertainties in parentheses are one standard deviation in units of the last quoted digit.}\label{tab:C13B_const}
\begin{tabular}{lll}
\hline
Constant     & $v=0$  & $v=1$  \\
\hline
$T_v$            & 25713.4438(29)   & 27504.4876(40)  \\ 
                 & 25713.4456(13)$^a$   & 27504.4859(21)$^a$  \\ 
$B_v$            & 12.56638(22)     & 11.09913(67)    \\ 
                 & 12.56652(11)$^a$     & 11.09953(20)$^a$    \\ 
$\gamma_v$$\times$10$^2$  & -2.700(74)       & -2.206(80)      \\ 
                   & -2.680(110)$^a$      & -2.260(110)$^a$     \\ 
$D_v$$\times$10$^3$     & 2.10830(400)         & 3.0660(290)     \\ 
                 & 2.10289(89)$^a$      & 3.0621(35)$^a$      \\ 
$H_v$$\times$10$^7$     & -3.540(270)      & -28.9(35)       \\ 
                 & -3.945(30)$^a$       & -29.57$^{a,b}$        \\ 
$L_v$$\times$10$^{10}$  & -1.31(60)        & 0.0             \\ 
                 &     0.0$^a$          & 0.0$^a$            \\ 
$\gamma_{Dv}$$\times$10$^5$  & 1.19(41)         & 0.0             \\ 
                      & 0.0$^a$              & 0.0$^a$            \\ 
\hline
\end{tabular}
\tablefoot{$^a$ Constants derived by \citet{Para1996}.}
\end{table*}
\begin{table*}
\caption{Constants (in cm$^{-1}$) for the  C$^2\Sigma^+$ state of $\rm ^{13}$CH. Uncertainties in parentheses are one standard deviation in units of the last quoted digit.  $^a$ Values from  \citet{Li1999}.}\label{tab:C13C_const}
\begin{tabular}{ll}
\hline
Constant     & $v=0$    \\
\hline                          
$T_v$            & 31791.8419(33)  \\ 
                 & 31791.8329(31)$^a$  \\ 
$B_v$            & 14.17064(20)    \\ 
                 & 14.17132(35)$^a$     \\ 
$\gamma_v$$\times$10$^2$  & 4.001(80)       \\ 
                   & 4.030(250)$^a$       \\ 
$D_v$$\times$10$^3$     & 1.5666(20)      \\ 
                 & 1.5728(27)$^a$       \\ 
$H_v$$\times$10$^8$     & 4.49(53)        \\ 
                 & 5.15(51)$^a$         \\ 
$\gamma_{Dv}$$\times$10$^5$  & 0.0       \\ 
                   & 2.03$^{a,b}$       \\ 
\hline
\end{tabular}
\tablefoot{$^b$ Constants kept fixed in \citet{Bembenek1997}. }
\end{table*}

\subsection{Equilibrium constants and partition function}
\subsubsection{Equilibrium constants}
In Table~\ref{tab:eqconst} we present equilibrium molecular constants (in $\rm cm^{-1}$) for the four lowest doublet electronic states of $\rm ^{12}$CH, including the $a^4\Sigma^-$ state for which the constants are taken from \citet{Kalemos1999} and \citet{Lie1973}. These constants were used to compute the potential curves and the transition-moment matrix elements (see Sect.~\ref{sec:methodo}). Moreover, our final results yield a dissociation energy for the CH molecule of 27960$\pm$ 10 cm$^{-1}$ ($= 3.4666\pm0.0013$\,eV) in agreement with the work of \citet{Kumar1998}.  

\begin{table*}
\caption{Equilibrium constants for $\rm ^{12}$CH}\label{tab:eqconst}
\begin{tabular}{lccccc}
\hline
Constant        &  X$^2\Pi$  &  $a^4\Sigma^-$ & A$^2\Delta$    & B$^2\Sigma^-$ & C$^2\Sigma^+$ \\ 
\hline                               
D$\rm_e (cm^{-1})$    & 29374.0  &   23349.6 & 16210.26 &   4731.57     &     7549.36             \\
D$\rm^0_{0} (cm^{-1})$ & 27959.64 &           &        &               &                 \\ 
$r_e ({\rm \AA})$           & 1.11981   & 1.0892    &   1.10366       &  1.16370      &  1.12777             \\ 
\hline                               
$T_e$           &   0.0          & 6024.40$^{a}$   &  23148.7375  &   24642.425            & 31809.6428               \\
$\omega_e$      &  2860.9238 & 3090.9$^{a}$    &  2923.3763   &   2255.07            &   2853.1724              \\
$\omega_ex_e$   &  64.58180  & 102.17$^{a}$    &  90.2    &    223.2$^c$       &   120.8564             \\
$\omega_ey_e$   &   0.407275   &           &  -0.1917    &               &                \\
$\omega_ez_e$   &  -0.01980  &           &  -0.3892     &               &                \\
\hline                                                                                 
$B_e$           &   14.45924  &  15.364$^b$   &   14.88536   &   13.3818     &   14.56066             \\
$\alpha_e$      &  0.53502   &  0.553$^b$     &   -0.591798  &   1.4828      &   0.56653             \\
$\gamma_e$      &   0.001993   &           &   -0.047110  &               &   -0.08659            \\
$\epsilon_e$    &  0.000214    &           &   0.00782    &               &                \\
\hline
\end{tabular}
\tablefoot{$^a$ Constants taken from \citet{Kalemos1999}.$^b$ Constants taken from \citet{Lie1973}. $^c$value fixed according to \citet{Lie1973}.}
\end{table*}

\subsubsection{Partition function}
\begin{figure}
\includegraphics[width=7cm,angle=-90]{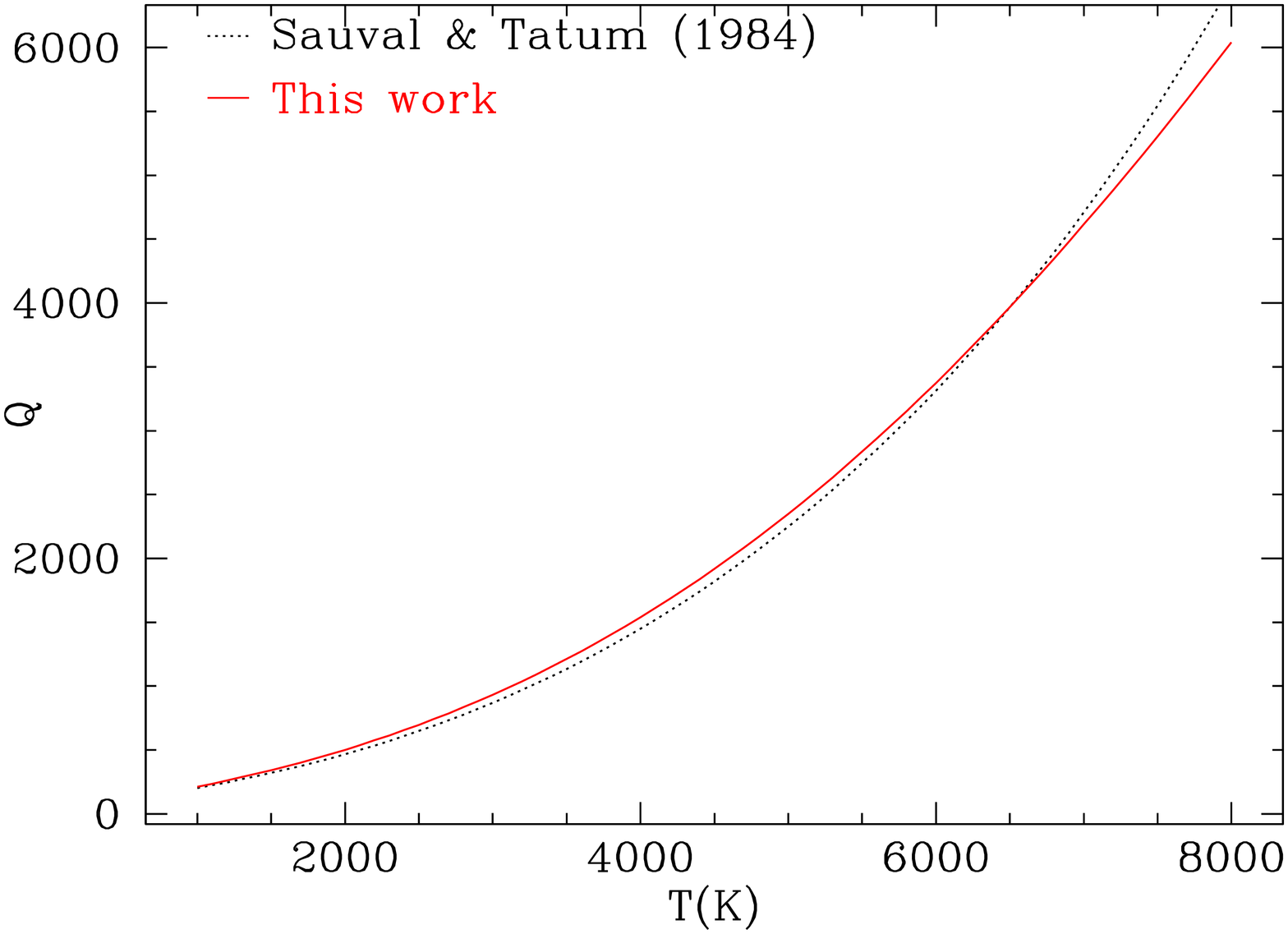}
\caption{Our new partition function compared to the work of \citet{Sauval1984}. The discrepancy is on average 3\%. }\label{Fig:partf} 
\end{figure}

Using our calculated levels up to the dissociation energy limit, we compute the partition function $Q$ following \citet{Tatum1966},
\begin{eqnarray}
Q(T) & = & \sum\limits_{i=0}^{E_i \le D^0_0} g_i e^{-\frac{E_i}{k.T}}, 
\end{eqnarray}
where $T$ is the temperature, $g_i=2J +1$ is the multiplicity of a single $\Lambda$-doubling rotational level of energy $E_i$, and $k$ is the Boltzmann constant.
For the $X^2\Pi$, $A^2\Delta$, $B^2\Sigma^-$, and $C^2\Sigma^+$ states we used the constants derived in this work. We also included the low-lying $a^4\Sigma^-$ state with constants from \citet{Lie1973} and \citet{Kalemos1999}, important for the accuracy of the partition function.  Following the formula $ln(Q)=\sum_0^2 a_i ln(T)^i$, the derived coefficients for our new partition function are $a_0 = 11.00608809, a_1=-2.68892131$ and $a_2=0.27091654$. This updated partition function differs from  that of \citet{Sauval1984} by 6 \% at T= 4000K, and 2 \% at 6000K (see Fig.~\ref{Fig:partf}).

\subsection{linelist format}
Our linelist includes the A-X, B-X, and C-X electronic transitions  up to the last observed vibrational levels (i.e. $v\le5$ for the X and A states, $v\le1$ for the B state, and $v\le2$ for the C state), with rotational levels up to the dissociation energy as listed in Table~\ref{tab:eqconst}.  Concerning the $\rm ^{13}$CH linelists, for the vibrational levels that could not be observed  but that exist for $\rm ^{12}$CH ( i.e. $X^2\Delta$ $v=4$ and 5, $A^2\Delta$ $v=4$ and 5, and C$^2\Sigma^+$ $v=1$ and 2), the Born-Oppenheimer approximation from $\rm ^{12}$CH has been used to determine the transitions applying the \citet{Jorgensen1994} formulas (3), (4), and (5). The X-X ro-vibrational transitions up to dissociation are also included for both isotopologues.
 
The linelist is available on http://www.astro.ulb.ac.be/$\sim$spectrotools/, and will be made available through the VALD database \citep{Kupka1999}. An example of the default output format is presented  in Table~\ref{tab:linelist}. The database offers various output formats and unit conversions. 
The gf value is defined as in \citet{Larsson1983}, 
\begin{eqnarray}
gf=(2 J'' + 1)\times f_{J''J'} &= & \frac{m_e \epsilon_0 c}{2\pi e^2 \nu^2} (2 J' + 1) A_{J''J'},
\end{eqnarray}
$f_{J'J''}$ being the absorption oscillator strength and $\nu$ the line position in cm$^{-1}$. 

Additionally, in our final linelist, we have included radiative and predissociation broadening (\rm $\Gamma_{\rm rad}$), using the formula 
\begin{eqnarray}
 \Gamma_{\rm rad}&=&1/\tau^{\rm rad}_{v'J'} + 1/\tau^{\rm predissoc}_{v'J'},
\end{eqnarray}
where $\tau^{\rm rad}$ is the radiative lifetime and $\tau^{\rm predissoc}$ is the predissociation lifetime of the level in seconds. This broadening includes both the radiative lifetime and the predissociation lifetime as measured by \citet{Luque1996_AX}, \citet{Luque1996_BX} and \citet{Brzozowski1976} for, respectively, the A, B, and C state.

\begin{landscape}
\begin{table*}
{\tiny
\caption{Sample of the linelist {\it(the complete list as well as the other electronic and ro-vibrational transitions are available electronically)}. }
\begin{tabular}{cccccccccccccccccc}
\hline
 $\lambda_{air}$(${\rm \AA}$)  &   gf    &       E$\rm _{low}(cm^{-1})$ & vl & J & Nl & sl  &  E$\rm _{up}(cm^{-1}$) &   vu & Ju & Nu & su  & $\Gamma_{rad}$  &   mol & trans & br & o-c(cm$^{-1}$) & reference \\
\hline
   4300.3005 & 0.1170117E-01  &       14.27597 &   0  & 0.5  &  0 &  e  &     23261.91647  &  0  & 1.5  &  0  & e  &    0.328588E+07 &  'CH  & A X  &  R  & -0.05040  & Bernath et al. (1991)' \\
   3847.4470 & 0.1054307E-04  &       14.27597 &   0  & 0.5  &  0 &  e  &     25998.16801  &  1  & 1.5  &  0  & e  &    0.311591E+07 &  'CH  & A X  &  R  &  0.00000  &  \\
   3504.5093 & 0.6155907E-05  &       14.27597 &   0  & 0.5  &  0 &  e  &     28540.78108  &  2  & 1.5  &  0  & e  &    0.609882E+07 &  'CH  & A X  &  R  &  0.00000  & \\
   3240.0300 & 0.1287390E-06  &       14.27597 &   0  & 0.5  &  0 &  e  &     30869.28429  &  3  & 1.5  &  0  & e  &    0.142829E+07 &  'CH  & A X  &  R  &  0.00000  & \\
   3034.9224 & 0.6833893E-08  &       14.27597 &   0  & 0.5  &  0 &  e  &     32954.46056  &  4  & 1.5  &  0  & e  &    0.116662E+07 &  'CH  & A X  &  R  &  0.00000  & \\
   2877.3544 & 0.1802274E-07  &       14.27597 &   0  & 0.5  &  0 &  e  &     34758.23015  &  5  & 1.5  &  0  & e  &    0.898224E+06 &  'CH  & A X  &  R  &  0.00000  & \\
   4872.9244 & 0.3025494E-03  &     2746.09918 &   1  & 0.5  &  1 &  e  &     23261.91647  &  0  & 1.5  &  1  & e  &    0.328588E+07 &  'CH  & A X  &  R  &  0.01051  & Zachwieja et al. (1995)' \\
   4299.4823 & 0.1028162E-01  &     2746.09918 &   1  & 0.5  &  1 &  e  &     25998.16801  &  1  & 1.5  &  1  & e  &    0.311591E+07 &  'CH  & A X  &  R  & -0.03690  & Bernath et al. (1991)' \\
\hline 
\end{tabular}
\label{tab:linelist}
}
\end{table*}
\end{landscape}

%\onltab{1}{}
%\onltab{2}{}
%\onltab{3}{}
%\onltab{4}{}
\section{Conclusion}
We have compiled the most up-to-date and extensive list of the main CH transitions observed in stellar spectra (ro-vibrational, A-X, B-X, and C-X). This list includes line positions as well as oscillator strengths and line broadening (radiative + predissociation). 
The combination of laboratory measurements and astronomical observations also allows the molecular data to be improved for this molecule. We have demonstrated that carefully selected stellar spectra can significantly improve molecular data despite their relatively poor quality compared to laboratory experiments or the solar spectrum. Although we were able to identify many new levels and we were able to assemble a large quantity of laboratory data, all levels of the CH molecules have not been identified. Therefore, the linelist we provide here is not complete, but is the most accurate exisiting for stellar spectroscopists.   
Other electronic transitions have been observed in some astrophysical contexts in the far UV \citep[D-X, E-X, and F-X;][]{Watson2001,Sheffer2007}. Laboratory data for these transitions can be found concerning the line positions and constants in \citet{Li1999_D} and \citet{Herzberg1969} for the D state, \citet{Watson2001} and \citet{Sheffer2007} for E and F, while \citet{Kalemos1999} provide theoretical values for all of them. The corresponding transition moments have also been calculated in \citet{vanDishoeck1987}. The D and F states are also known to be highly predissociative \citep{Herzberg1969}. Predissociation lifetimes for these states can be found in \citet{Metropoulos2000}. However, we did not include these transitions in our linelists because they are expected to be of minor importance for stellar spectra.
We also highlight the fact that predissociation is not only known to exist in CH, but also in other molecules of astrophysical interest (NH, H$_2$O, etc...) \citep[see][]{Kato1995}.

\begin{acknowledgements}
TM, SvE and AJ are  supported in part by an {\it Action de Recherche Concert\'ee from
the Direction g\'en\'erale de l'enseignement non obligatoire et de la
Recherche Scientifique -- Direction de la Recherche Scientifique --
Communaut\'e Fran\c{c}aise de Belgique} and by the F.R.S.-FNRS FRFC grant
2.4533.09. Some support was also provided by the NASA laboratory astrophysics program. NC acknowledges support by Sonderforschungsbereich SFB 881 "The Milky Way
System" (subprojects A2, A4, and A5) of the German Research Foundation(DFG). This work has used data from ESO program ID 69.D-0063(A).
\end{acknowledgements}

\appendix
\section{Atlas of predissociation lines}
\subsection{Predissociation lines in astronomical spectra}
In the course of a study of carbon-enhanced metal-poor (CEMP) stars \citep{Christlieb2002,MasseronPhD,Masseron2006,Masseron2010,Masseron2012}, we came across wide and shallow lines spread in the wavelength ranges $\lambda 3660 - 3770$~\AA\ and $\lambda 3920 - 4130$~\AA\,(Fig.~\ref{fig:BX}). Realising that these ranges matched those of the B-X system of the CH molecule, it soon became clear that these wide and shallow lines are in fact predissociation lines. It prompted the present work on the CH molecular linelist.

In molecules, different types of transitions may occur. If the energy provided to the molecule is higher than that needed to separate the atomic components, then dissociation occurs and no lines are formed. If both the upper and lower states of the transition are below the dissociation limit, the classical quantized transition occurs and a line forms. However, if a bound transition involves a level having a nearby unbound level, an internal conversion may occur, leaving the transition in the so-called predissociation state. Because the unbound level has a finite lifetime associated with the molecular dissociation, the lines associated with such transitions appear broadened beyond the natural broadening \citep{Herzberg1950} once the predissociation lifetime becomes shorter than the radiative lifetime. There are various situations leading to predissociation. The unbound level can be of the same or a different electronic state, but certain selection rules must be fulfilled \citep[see][]{Herzberg1950}. Furthermore, the predissociation may be internal to a molecule but can also be induced by external factors (e.g. magnetic fields).

\begin{figure}[h!]
\includegraphics[width=7cm]{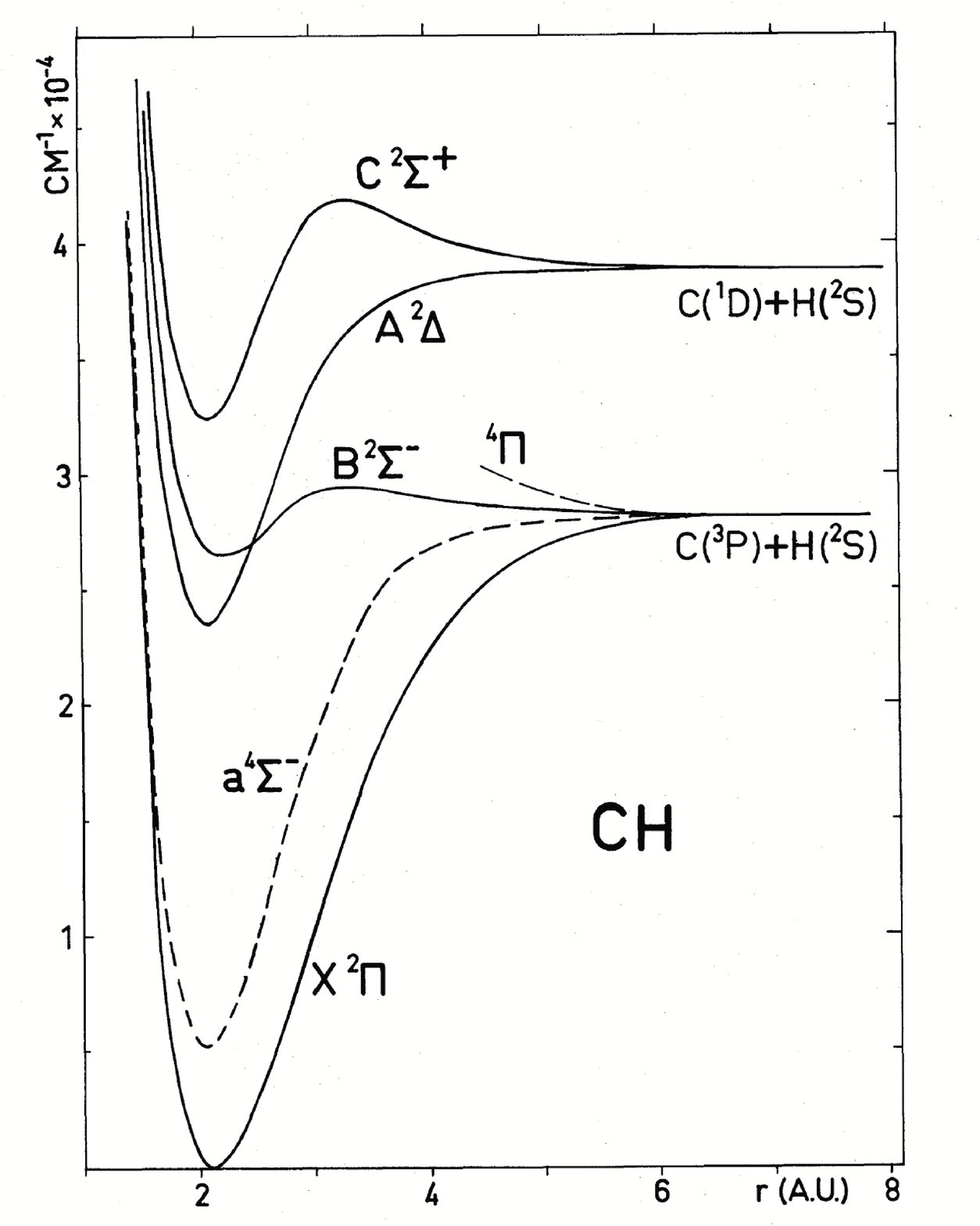}
\caption{Potential curves of the lowest excited electronic states in CH \citep[extracted from][]{Brzozowski1976}}\label{fig:CH-potcurve}
\end{figure}

For the CH molecule,  predissociation is expected for the A, B, C, and D states. Predissociation was first observed in the laboratory by \citet{Shidei1936} and confirmed by \citet{Herzberg1969}, and \citet{Ubachs1986} and studied theoretically in more detail by \citet{Elander1973}, \citet{Brooks1974}, \citet{Elander1979}, and \citet{vanDishoeck1987}.
For the A state, \citet{Brzozowski1976} noticed that the line structure disappears for energies below the dissociation limit, but reappears again at higher energies. They attribute this predissociation to perturbation of the X$^2\Pi$ ground state and from the B$^2\Sigma^-$ state.
\citet{Brzozowski1976} and  \citet{Elander1979} show that the B-state has a potential barrier above the dissociation limit (see Fig.~\ref{fig:CH-potcurve}), so that its predissociation occurs by tunneling through the barrier by quasibound levels. 
Although a similar bump in the potential curve of the C state is observed, \citet{Brzozowski1976} and \citet{Elander1973} first suspected the a$^4\Sigma$ to be responsible for the perturbation of the C state leading to its predissociation. Nevertheless, more recent calculations of the C state by \citet{vanDishoeck1987} and observations by \citet{Ubachs1986} show that the predissociation of the C state is instead due to interaction with the B state. 
Finally, \citet{vanDishoeck1987} also predicted the D state to be entirely predissociated. 

Although the predissociation phenomenon has been extensively observed in the laboratory, predissociation lines have rarely been identified in astronomical objects: NH$_3$ and NH$_2$ predissociation lines have been observed in Halley's comet \citep{Fink1992}. Predissociation lines are often discussed in relation with diffuse interstellar bands \citep[DIBs; see][]{Herbig1975,Herbig1995,Fulara2000}. Hubble Space Telescope observations of \citet{Watson2001} assign the $\sim1370$~\AA\, DIBs to the CH molecule.  Later on, \citet{Sheffer2007} confirmed this finding and extended it to the 1271, 1549, and 1694~\AA\,DIBs.
Carbon-rich stars present a large number of molecular features, and notably of CH, including predissociation lines. Because broadening in stellar spectra is large compared to laboratory spectra, only B-X predissociation lines are detectable. However, the thermodynamical conditions in their atmospheres allow us to observe a large number of those lines that we list in the following atlas.

\subsection{Atlas}
\label{Sect:atlas}

We present here an atlas of predissociation lines in carbon-rich stars (see Table~\ref{Tab:carbonstars} for the stellar sample), the Sun, and Arcturus. The catalog (Figs. A.3 -- A.10) clearly reveals the peculiar shapes that may be encountered in spectra, especially in carbon-rich stars where they are the most conspicuous, but also in normal stars like the Sun. 

\begin{table*}
\caption{Stellar parameters of the stars presented in the Atlas.}
\begin{tabular}{l|c|c|c|c|c}
\hline
Star          &  $\rm T_{eff}$   & log(g)  & [Fe/H]  &  [C/Fe] & reference \\
\hline
HD~187861     &  4600  &  1.7  &   -2.4   &  2.1   &   \citet{Masseron2010}          \\
CS~22942-019  &  5100  &  2.5  &   -2.5   &  2.2   &    \citet{Masseron2010}           \\
HE~1419-1324  &  5600  &  3.2  &  -2.1    &  1.2   &    \citet{Masseron2010}           \\
Arcturus      & 4300   &  1.5  &  -0.5    &  0.0   &    \citet{Decin2003} \\
Sun           & 5777   &  4.44 &  0.0     &   0.0  &             \\
\hline  
 \end{tabular}\label{Tab:carbonstars}                                                                                  
\end{table*}

\subsection {The Bond-Neff depression}

In this section we evaluate the impact of the CH bands on low-resolution spectra and photometry, especially in the context of the Bond-Neff depression \citep{Bond1969} observed in barium stars \citep{Bidelman1951}, which are G-K giants enriched in carbon and heavy elements predominantly produced in the s-process nucleosynthesis \citep[see][for a recent review]{Kappeler2011}. 

\begin{figure}
\includegraphics[width=7cm,angle=-90]{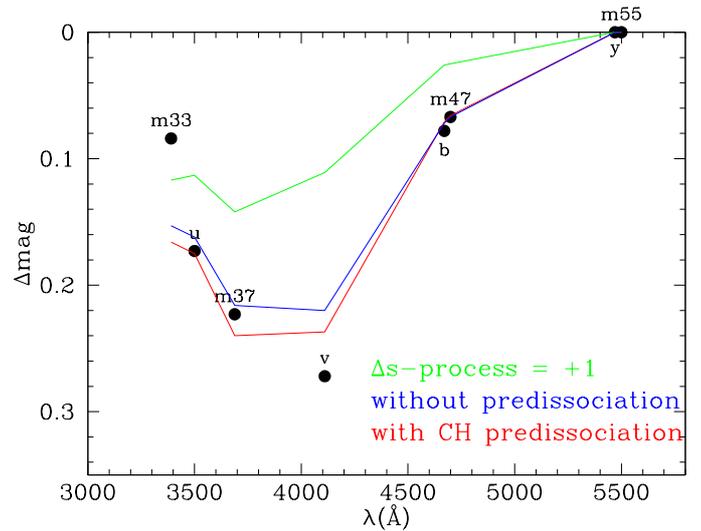}
\caption{Drop in magnitudes in the blue-yellow spectra of the barium star $\zeta$~Cap compared to the normal giant o~Uma due to blanketing by heavy elements and by CH and CN. The black dots correspond to different photometric bands, as taken from \citet{Bond1969}, and the continuous lines correspond to the synthetic spectra with s-process elements increased by +1~ dex (green line) and with CH added with (red line) or without (blue line) predissociation lines. While similar atmospheric parameters were used, carbon and s-process element overabundances of 0.5 dex and 1 dex were adopted, respectively.}\label{Fig:BondNeff} 
\end{figure}

Although the Bond-Neff depression was first assigned to CN and CH molecules, the depression was not satisfactorily accounted for with the old linelists. \citet{McWilliam1984} therefore suggested a contribution from blanketing by heavy elements. By computing synthetic spectra for two among the barium stars originally considered by \citet{Bond1969}, we show in Fig.~\ref{Fig:BondNeff} that there is now a very good agreement between predicted and observed colours, thanks to our complete molecular and atomic linelists. As demonstrated by this figure, both heavy-element and CH blanketing (including CH-predissociation lines) are necessary to reproduce the Bond-Neff depression. 

\begin{figure*}[h!]
\includegraphics[height=25cm]{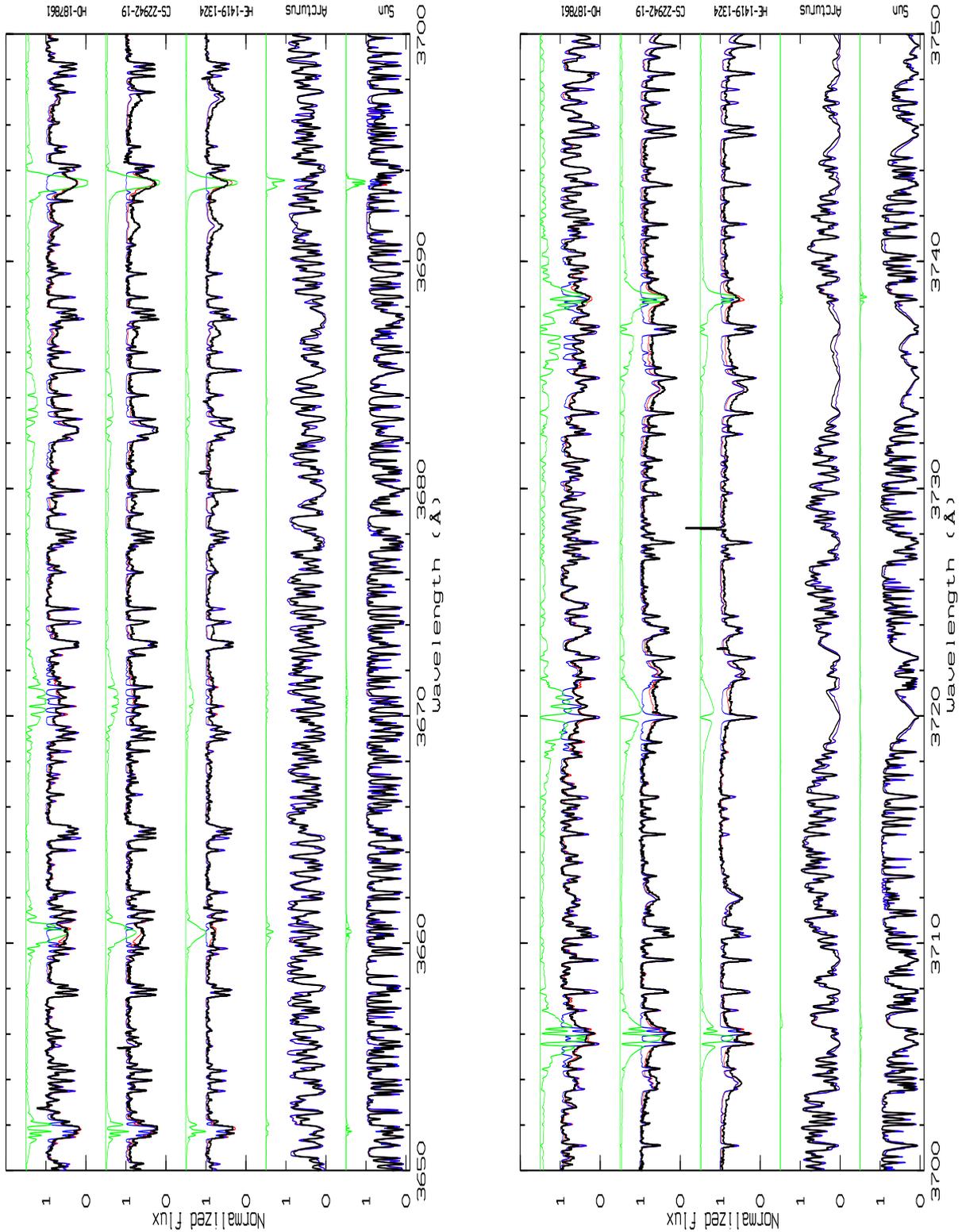}
\caption{Atlas of predissociation lines for a sample of stars. The black line shows the observations while the red lines shows the synthetic fit with all atomic and molecular lines included, the blue line shows the synthetic fit without CH predissociation lines and the green line shows the contribution of the CH predissociation lines alone. We note that the continuum has been arbitrarily placed at 1.5 for the synthetic fit. See Table~\ref{Tab:carbonstars} for a list of the basic properties of the displayed stars.}
\label{fig:atlas1}
\end{figure*}
\begin{figure*}[h!]
\includegraphics[height=25cm]{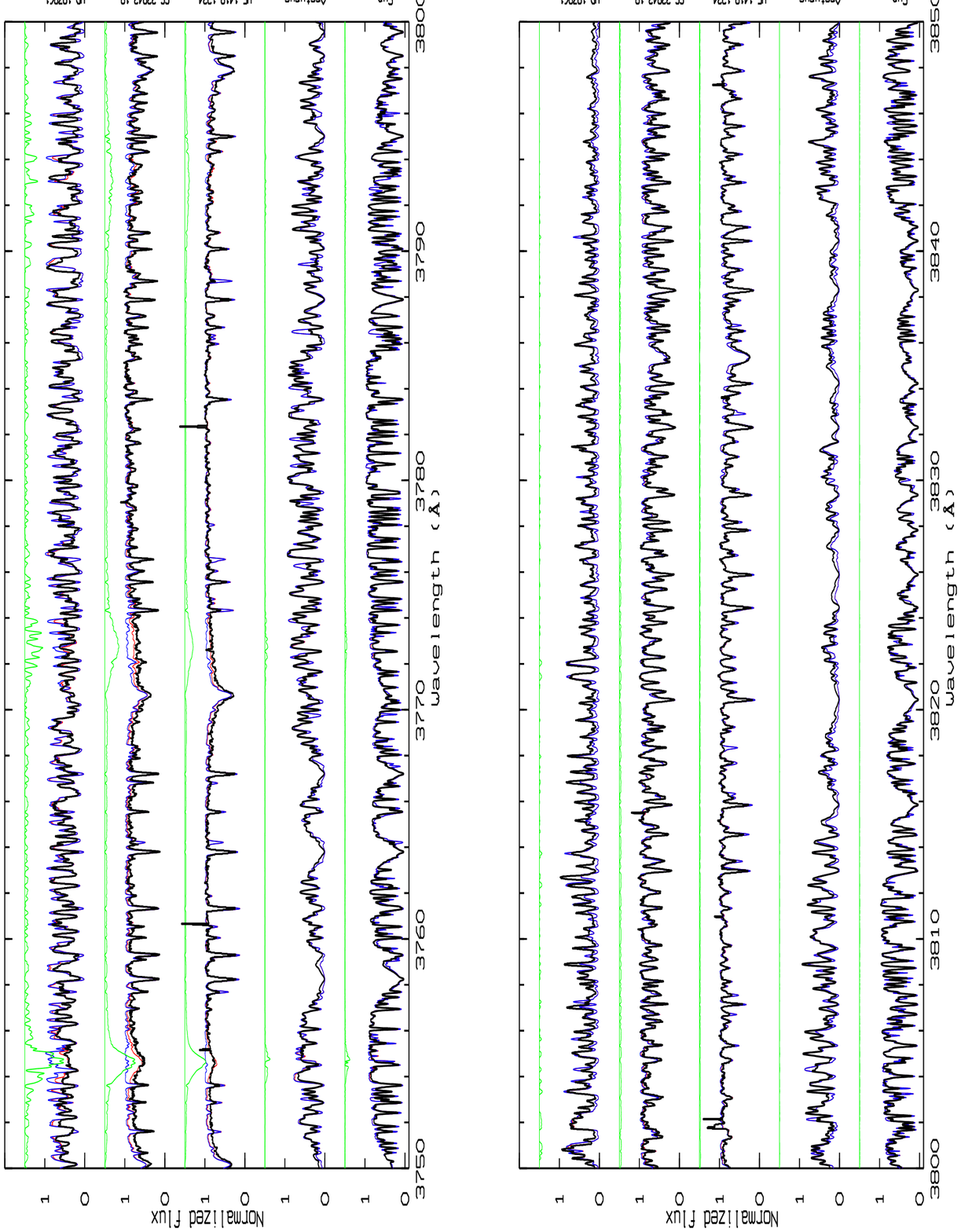}
\caption{Atlas of predissociation lines (continued).}
\label{fig:atlas2}
\end{figure*}
\begin{figure*}[h!]
\includegraphics[height=25cm]{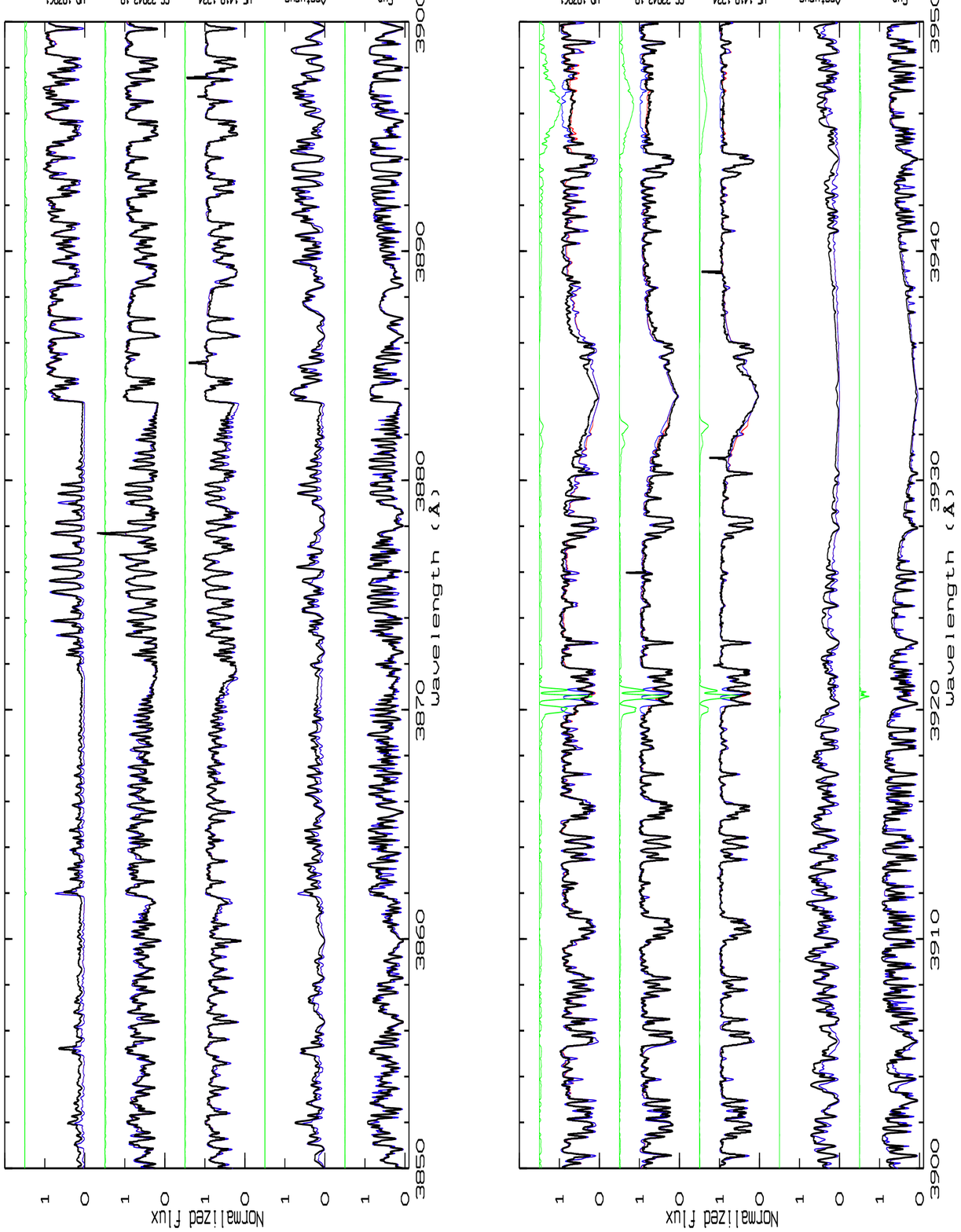}
\caption{Atlas of predissociation lines (continued).}
\label{fig:atlas3}
\end{figure*}
\begin{figure*}[h!]
\includegraphics[height=25cm]{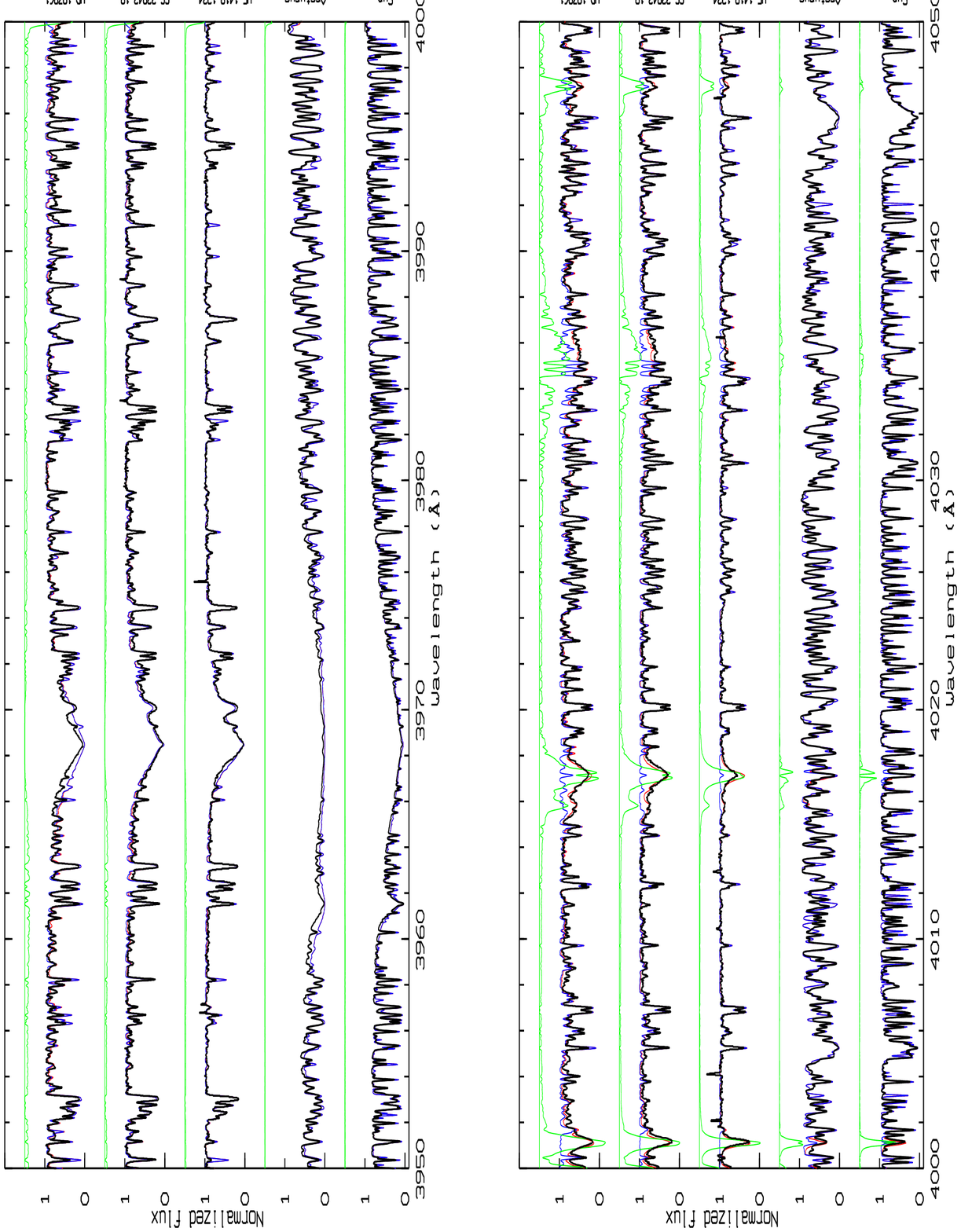}
\caption{Atlas of predissociation lines (continued).}
\label{fig:atlas4}
\end{figure*}
\begin{figure*}[h!]
\includegraphics[height=25cm]{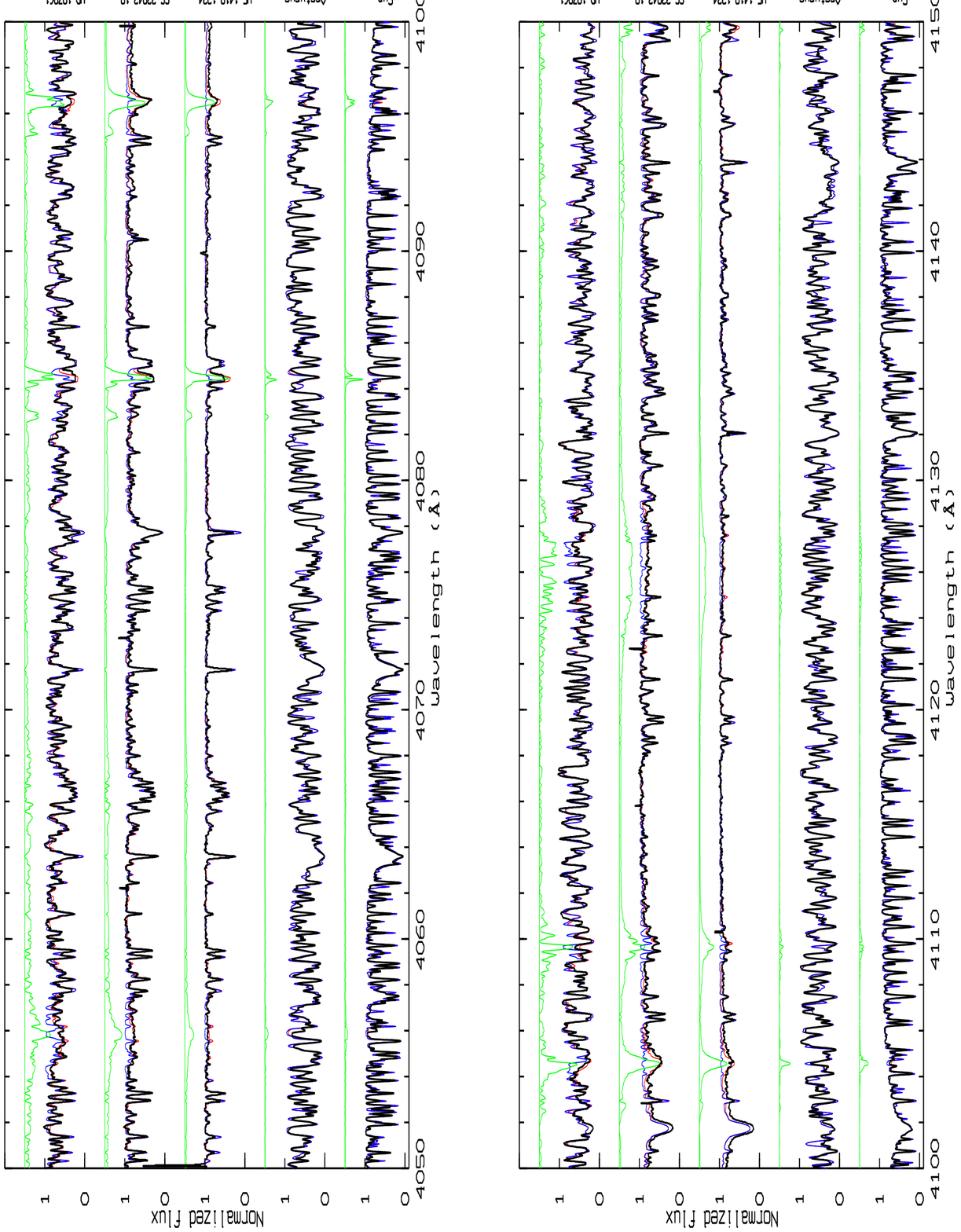}
\caption{Atlas of predissociation lines (continued).}
\label{fig:atlas5}
\end{figure*}
\begin{figure*}[h!]
\includegraphics[height=25cm]{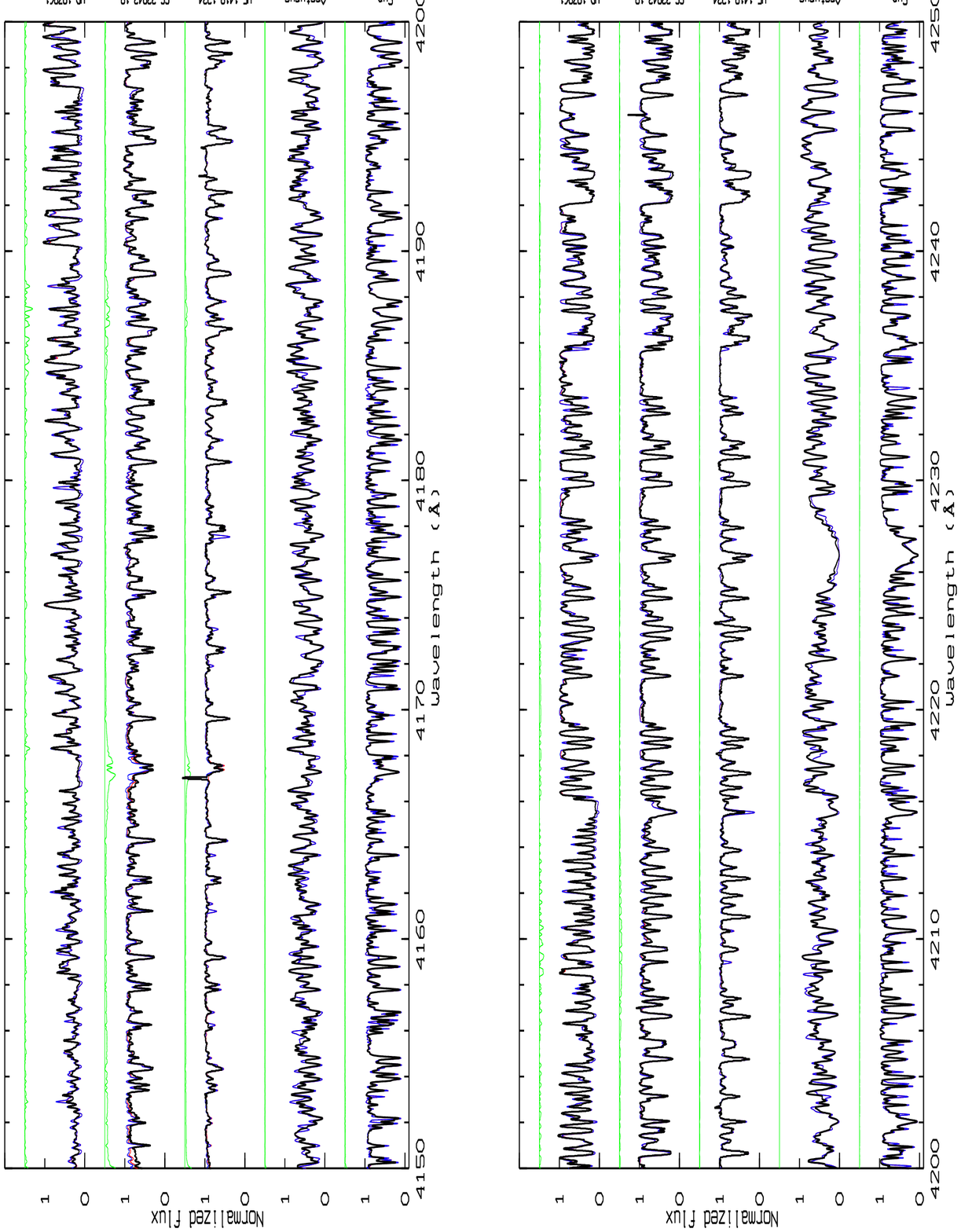}
\caption{Atlas of predissociation lines (continued).}
\label{fig:atlas6}
\end{figure*}
\begin{figure*}[h!]
\includegraphics[height=25cm]{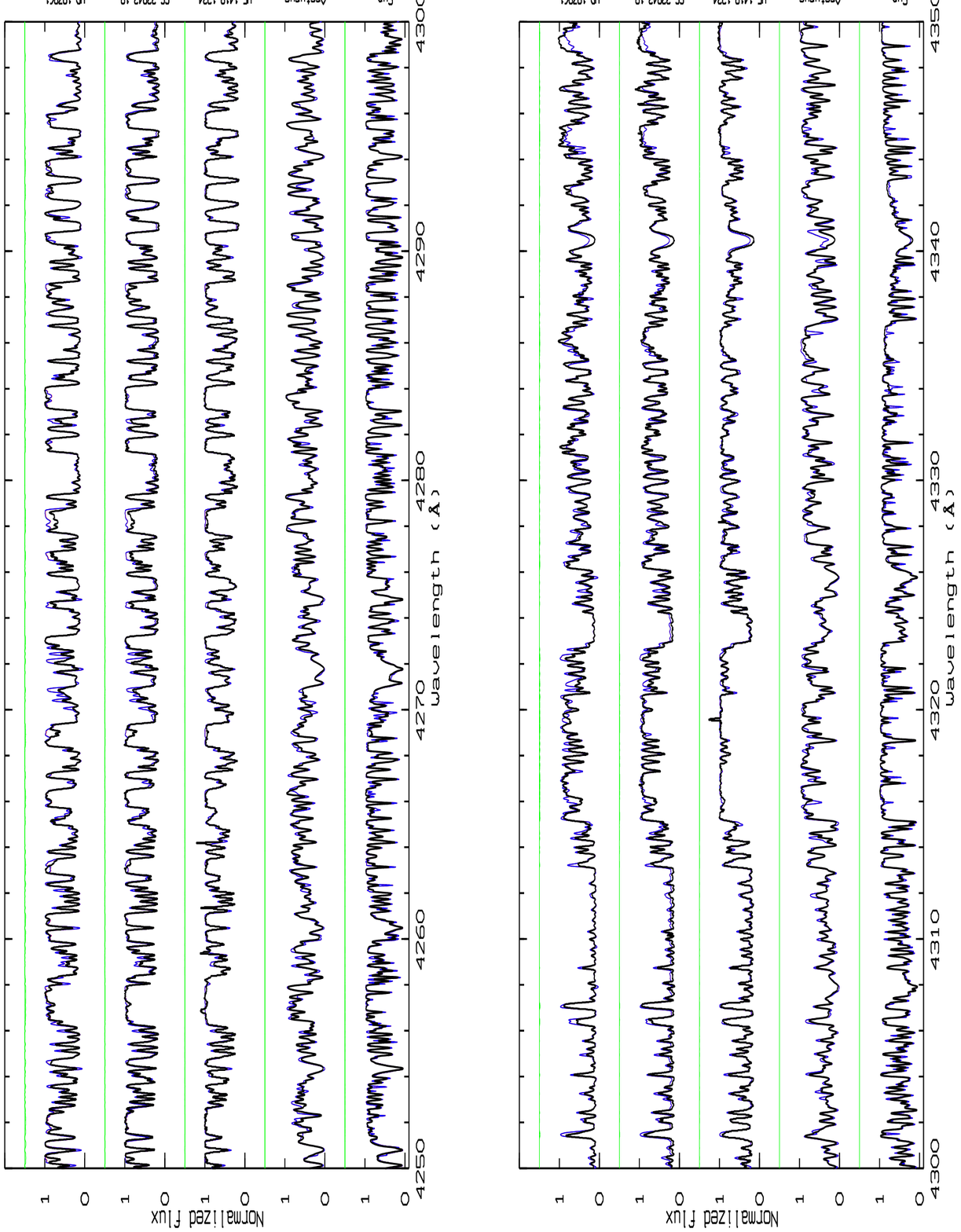}
\caption{Atlas of predissociation lines (continued).}
\label{fig:atlas7}
\end{figure*}
\begin{figure*}[h!]
\includegraphics[height=25cm]{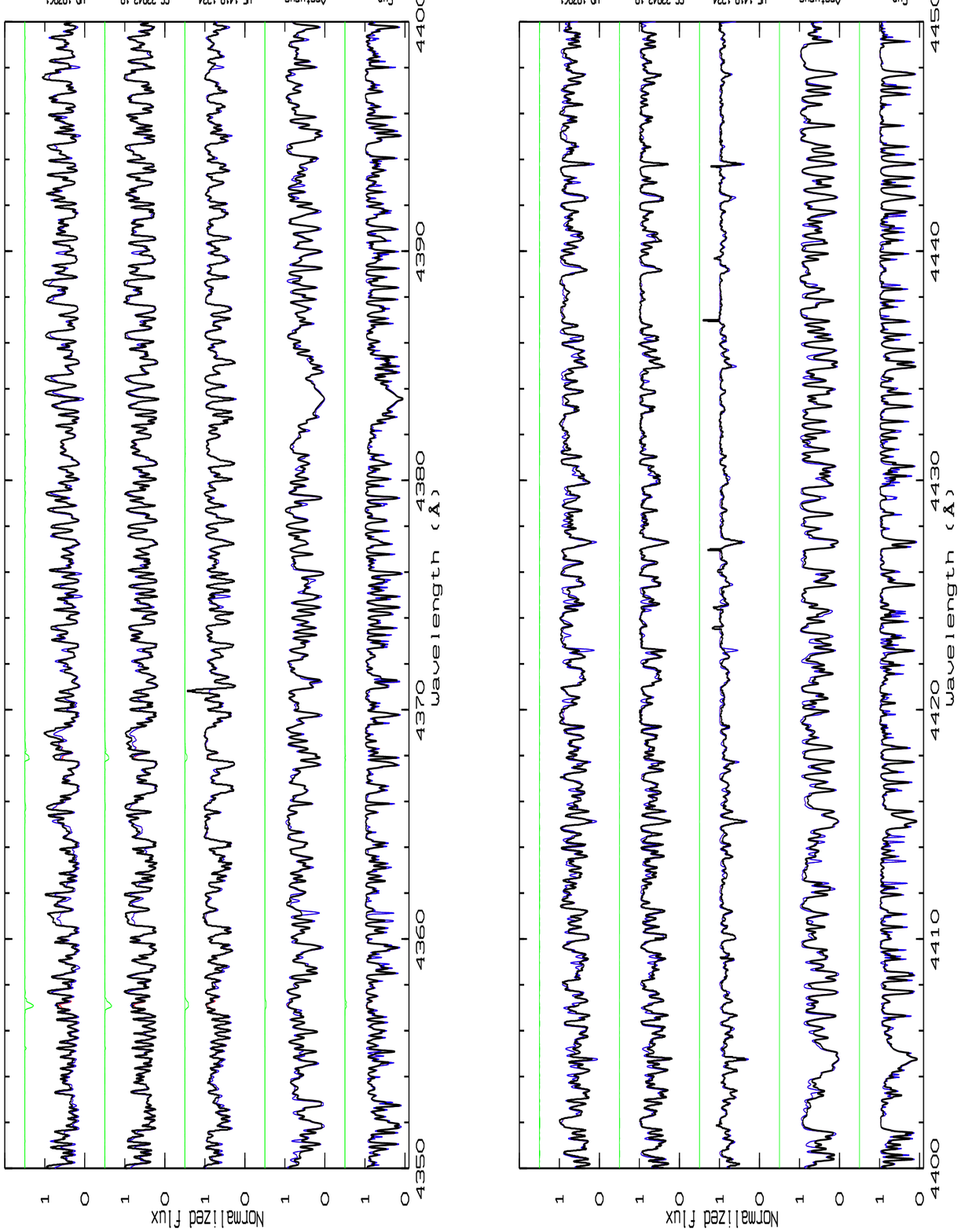}
\caption{Atlas of predissociation lines (continued).}
\label{fig:atlas8}
\end{figure*}

\bibliographystyle{aa}
\bibliography{predissoc}

\begin{thebibliography}{68}
\expandafter\ifx\csname natexlab\endcsname\relax\def\natexlab#1{#1}\fi

\bibitem[{{Alvarez} \& {Plez}(1998)}]{Alvarez1998}
{Alvarez}, R. \& {Plez}, B. 1998, \aap, 330, 1109

\bibitem[{{Asplund}(2005)}]{2005ARA&A..43..481A}
{Asplund}, M. 2005, \araa, 43, 481

\bibitem[{{Bembenek} {et~al.}(1997){Bembenek}, {Ke}, \& {Rytel}}]{Bembenek1997}
{Bembenek}, Z., {Ke}, R., \& {Rytel}, M. 1997, Journal of Molecular
  Spectroscopy, 183, 1

\bibitem[{{Bernath} {et~al.}(2005){Bernath}, {McElroy}, {Abrams}, {Boone},
  {Butler}, {Camy-Peyret}, {Carleer}, {Clerbaux}, {Coheur}, {Colin}, {DeCola},
  {De Mazière}, {Drummond}, {Dufour}, {Evans}, {Fast}, {Fussen}, {Gilbert},
  {Jennings}, {Llewellyn}, {Lowe}, {Mahieu}, {McConnell}, {McHugh}, {McLeod},
  {Michaud}, {Midwinter}, {Nassar}, {Nichitiu}, {Nowlan}, {Rinsland}, {Rochon},
  {Rowlands}, {Semeniuk}, {Simon}, {Skelton}, {Sloan}, {Soucy}, {Strong},
  {Tremblay}, {Turnbull}, {Walker}, {Walkty}, {Wardle}, {Wehrle}, {Zander}, \&
  {Zou}}]{Bernath2005}
{Bernath}, P., {McElroy}, C., {Abrams}, M., {et~al.} 2005, Geophys. Res. Lett.
  32, L15S01, See also http://www.ace.uwaterloo.ca/

\bibitem[{{Bernath}(2009)}]{Bernath2009}
{Bernath}, P.~F. 2009, International Reviews in Physical Chemistry, 28, 681

\bibitem[{{Bernath} {et~al.}(1991){Bernath}, {Brazier}, {Olsen}, {Hailey},
  {Fernando}, {Woods}, \& {Hardwick}}]{Bernath1991}
{Bernath}, P.~F., {Brazier}, C.~R., {Olsen}, T., {et~al.} 1991, Journal of
  Molecular Spectroscopy, 147, 16

\bibitem[{{Bidelman} \& {Keenan}(1951)}]{Bidelman1951}
{Bidelman}, W.~P. \& {Keenan}, P.~C. 1951, \apj, 114, 473

\bibitem[{{Bi{\'e}mont} {et~al.}(1999){Bi{\'e}mont}, {Palmeri}, \&
  {Quinet}}]{Biemont1999}
{Bi{\'e}mont}, E., {Palmeri}, P., \& {Quinet}, P. 1999, \apss, 269, 635

\bibitem[{{Bond} \& {Neff}(1969)}]{Bond1969}
{Bond}, H.~E. \& {Neff}, J.~S. 1969, \apj, 158, 1235

\bibitem[{{Brooks} \& {Smith}(1974)}]{Brooks1974}
{Brooks}, N.~H. \& {Smith}, W.~H. 1974, \apj, 194, 513

\bibitem[{{Brown} {et~al.}(1979){Brown}, {Colbourn}, {Watson}, \&
  {Wayne}}]{Brown1979}
{Brown}, J.~M., {Colbourn}, E.~A., {Watson}, J.~K.~G., \& {Wayne}, F.~D. 1979,
  Journal of Molecular Spectroscopy, 74, 294

\bibitem[{{Brzozowski} {et~al.}(1976){Brzozowski}, {Bunker}, {Elander}, \&
  {Erman}}]{Brzozowski1976}
{Brzozowski}, J., {Bunker}, P., {Elander}, N., \& {Erman}, P. 1976, \apj, 207,
  414

\bibitem[{{Christlieb} {et~al.}(2002){Christlieb}, {Bessell}, {Beers},
  {Gustafsson}, {Korn}, {Barklem}, {Karlsson}, {Mizuno-Wiedner}, \&
  {Rossi}}]{Christlieb2002}
{Christlieb}, N., {Bessell}, M.~S., {Beers}, T.~C., {et~al.} 2002, \nat, 419,
  904

\bibitem[{{Colin} \& {Bernath}(2010)}]{Colin2010}
{Colin}, R. \& {Bernath}, P.~F. 2010, Journal of Molecular Spectroscopy, 263,
  120

\bibitem[{{Decin} {et~al.}(2003){Decin}, {Vandenbussche}, {Waelkens}, {Decin},
  {Eriksson}, {Gustafsson}, {Plez}, \& {Sauval}}]{Decin2003}
{Decin}, L., {Vandenbussche}, B., {Waelkens}, C., {et~al.} 2003, \aap, 400, 709

\bibitem[{Elander {et~al.}(1979)Elander, Hehenberger, \& Bunker}]{Elander1979}
Elander, N., Hehenberger, M., \& Bunker, P. 1979, Physica Scripta, 20, 631

\bibitem[{{Elander} \& {Smith}(1973)}]{Elander1973}
{Elander}, N. \& {Smith}, W.~H. 1973, \apj, 184, 663

\bibitem[{{Fink} {et~al.}(1992){Fink}, {Combi}, \& {Disanti}}]{Fink1992}
{Fink}, U., {Combi}, M., \& {Disanti}, M.~A. 1992, in Asteroids, Comets,
  Meteors 1991, ed. {A.~W.~Harris \& E.~Bowell}, 187--189

\bibitem[{{Fulara} \& {Kre{\l}owski}(2000)}]{Fulara2000}
{Fulara}, J. \& {Kre{\l}owski}, J. 2000, \nar, 44, 581

\bibitem[{{Gustafsson} {et~al.}(2008){Gustafsson}, {Edvardsson}, {Eriksson},
  {J{\o}rgensen}, {Nordlund}, \& {Plez}}]{Gustafsson2008}
{Gustafsson}, B., {Edvardsson}, B., {Eriksson}, K., {et~al.} 2008, \aap, 486,
  951

\bibitem[{{Gustafsson} {et~al.}(2010){Gustafsson}, {Mel{\'e}ndez}, {Asplund},
  \& {Yong}}]{2010Ap&SS.328..185G}
{Gustafsson}, B., {Mel{\'e}ndez}, J., {Asplund}, M., \& {Yong}, D. 2010, \apss,
  328, 185

\bibitem[{{Heimer}(1932)}]{Heimer1932}
{Heimer}, T. 1932, Zeitschrift fur Physik, 78, 771

\bibitem[{{Herbig}(1975)}]{Herbig1975}
{Herbig}, G.~H. 1975, \apj, 196, 129

\bibitem[{{Herbig}(1995)}]{Herbig1995}
{Herbig}, G.~H. 1995, \araa, 33, 19

\bibitem[{{Herzberg}(1950)}]{Herzberg1950}
{Herzberg}, G. 1950, {Molecular spectra and molecular structure. Vol.1: Spectra
  of diatomic molecules}, ed. {Van Nostrand Reinhold}

\bibitem[{{Herzberg} \& {Johns}(1969)}]{Herzberg1969}
{Herzberg}, G. \& {Johns}, J.~W.~C. 1969, \apj, 158, 399

\bibitem[{Hettema \& Yarkony(1994)}]{Hettema1994}
Hettema, H. \& Yarkony, D.~R. 1994, Journal of Chemical Physics, 100, 8991

\bibitem[{{Heurlinger} \& {Hulthen}(1919)}]{Heurlinger1919}
{Heurlinger}, T. \& {Hulthen}, E. 1919, Z. Wiss. Photogr. Photophys.
  Photochem., 18, 241

\bibitem[{{Hill} {et~al.}(2002){Hill}, {Plez}, {Cayrel}, {Beers}, {Nordstr{\"
  o}m}, {Andersen}, {Spite}, {Spite}, {Barbuy}, {Bonifacio}, {Depagne},
  {Fran{\c c}ois}, \& {Primas}}]{Hill2002}
{Hill}, V., {Plez}, B., {Cayrel}, R., {et~al.} 2002, \aap, 387, 560

\bibitem[{{Jackson} {et~al.}(2008){Jackson}, {Zink}, {McCarthy}, {Perez}, \&
  {Brown}}]{Jackson2008}
{Jackson}, M., {Zink}, L.~R., {McCarthy}, M.~C., {Perez}, L., \& {Brown}, J.~M.
  2008, Journal of Molecular Spectroscopy, 247, 128

\bibitem[{{Jorgensen}(1994)}]{Jorgensen1994}
{Jorgensen}, U.~G. 1994, \aap, 284, 179

\bibitem[{{Jorgensen} {et~al.}(1996){Jorgensen}, {Larsson}, {Iwamae}, \&
  {Yu}}]{Jorgensen1996}
{Jorgensen}, U.~G., {Larsson}, M., {Iwamae}, A., \& {Yu}, B. 1996, \aap, 315,
  204

\bibitem[{Kalemos {et~al.}(1999)Kalemos, Mavridis, \&
  Metropoulos}]{Kalemos1999}
Kalemos, A., Mavridis, A., \& Metropoulos, A. 1999, Journal of Chemical
  Physics, 111, 9536

\bibitem[{{K{\"a}ppeler} {et~al.}(2011){K{\"a}ppeler}, {Gallino}, {Bisterzo},
  \& {Aoki}}]{Kappeler2011}
{K{\"a}ppeler}, F., {Gallino}, R., {Bisterzo}, S., \& {Aoki}, W. 2011, Reviews
  of Modern Physics, 83, 157

\bibitem[{Kato \& Baba(1995)}]{Kato1995}
Kato, H. \& Baba, M. 1995, Chemical Reviews, 95, 2311

\bibitem[{{Kepa}(1996)}]{Kepa1996}
{Kepa}, R. 1996, Journal of Molecular Spectroscopy, 178, 189

\bibitem[{{Kumar} {et~al.}(1998){Kumar}, {Hsiao}, {Hung}, \& {Lee}}]{Kumar1998}
{Kumar}, A., {Hsiao}, C.-C., {Hung}, W.-C., \& {Lee}, Y.-P. 1998, \jcp, 109,
  3824

\bibitem[{{Kupka} {et~al.}(1999){Kupka}, {Piskunov}, {Ryabchikova}, {Stempels},
  \& {Weiss}}]{Kupka1999}
{Kupka}, F., {Piskunov}, N., {Ryabchikova}, T.~A., {Stempels}, H.~C., \&
  {Weiss}, W.~W. 1999, \aaps, 138, 119

\bibitem[{{Larsson}(1983)}]{Larsson1983}
{Larsson}, M. 1983, \aap, 128, 291

\bibitem[{{Li} {et~al.}(2012){Li}, {Harrison}, {Ram}, {Western}, \&
  {Bernath}}]{Li2012}
{Li}, G., {Harrison}, J.~J., {Ram}, R.~S., {Western}, C.~M., \& {Bernath},
  P.~F. 2012, \jqsrt, 113, 67

\bibitem[{Li {et~al.}(1999)Li, Kumar, Hsiao, \& Lee}]{Li1999}
Li, X., Kumar, A., Hsiao, C.-C., \& Lee, Y.-P. 1999, The Journal of Physical
  Chemistry A, 103, 6162

\bibitem[{{Li} \& {Lee}(1999)}]{Li1999_D}
{Li}, X. \& {Lee}, Y.-P. 1999, \jcp, 111, 4942

\bibitem[{{Lie} {et~al.}(1973){Lie}, {Hinze}, \& {Liu}}]{Lie1973}
{Lie}, G.~C., {Hinze}, J., \& {Liu}, B. 1973, \jcp, 59, 1887

\bibitem[{{Luque} \& {Crosley}(1999)}]{LIFBASE1999}
{Luque}, J. \& {Crosley}, D. 1999, in SRI International Report MP, Vol.~99, 009

\bibitem[{{Luque} \& {Crosley}(1996{\natexlab{a}})}]{Luque1996_AX}
{Luque}, J. \& {Crosley}, D.~R. 1996{\natexlab{a}}, \jcp, 104, 2146

\bibitem[{{Luque} \& {Crosley}(1996{\natexlab{b}})}]{Luque1996_BX}
{Luque}, J. \& {Crosley}, D.~R. 1996{\natexlab{b}}, \jcp, 104, 3907

\bibitem[{{Masseron}(2006)}]{MasseronPhD}
{Masseron}, T. 2006, PhD thesis, Observatoire de Paris, France

\bibitem[{{Masseron} {et~al.}(2012){Masseron}, {Johnson}, {Lucatello},
  {Karakas}, {Plez}, {Beers}, \& {Christlieb}}]{Masseron2012}
{Masseron}, T., {Johnson}, J.~A., {Lucatello}, S., {et~al.} 2012, \apj, 751, 14

\bibitem[{{Masseron} {et~al.}(2010){Masseron}, {Johnson}, {Plez}, {van Eck},
  {Primas}, {Goriely}, \& {Jorissen}}]{Masseron2010}
{Masseron}, T., {Johnson}, J.~A., {Plez}, B., {et~al.} 2010, \aap, 509, A93

\bibitem[{{Masseron} {et~al.}(2006){Masseron}, {van Eck}, {Famaey}, {Goriely},
  {Plez}, {Siess}, {Beers}, {Primas}, \& {Jorissen}}]{Masseron2006}
{Masseron}, T., {van Eck}, S., {Famaey}, B., {et~al.} 2006, \aap, 455, 1059

\bibitem[{{McWilliam} \& {Smith}(1984)}]{McWilliam1984}
{McWilliam}, A. \& {Smith}, V.~V. 1984, in Bulletin of the American
  Astronomical Society, Vol.~16, Bulletin of the American Astronomical Society,
  973

\bibitem[{Metropoulos \& Mavridis(2000)}]{Metropoulos2000}
Metropoulos, A. \& Mavridis, A. 2000, Chemical Physics Letters, 331, 89

\bibitem[{{Para}(1996)}]{Para1996}
{Para}, A. 1996, Journal of Physics B Atomic Molecular Physics, 29, 5765

\bibitem[{{Plez}(2008)}]{2008PhST..133a4003P}
{Plez}, B. 2008, Physica Scripta Volume T, 133, 014003

\bibitem[{{Plez}(2012)}]{2012ascl.soft05004P}
{Plez}, B. 2012, {Turbospectrum: Code for spectral synthesis}, astrophysics
  Source Code Library

\bibitem[{{Sauval} \& {Tatum}(1984)}]{Sauval1984}
{Sauval}, A.~J. \& {Tatum}, J.~B. 1984, \apjs, 56, 193

\bibitem[{{Sheffer} \& {Federman}(2007)}]{Sheffer2007}
{Sheffer}, Y. \& {Federman}, S.~R. 2007, \apj, 659, 1352

\bibitem[{{Shelyag} {et~al.}(2004){Shelyag}, {Sch{\"u}ssler}, {Solanki},
  {Berdyugina}, \& {V{\"o}gler}}]{2004A&A...427..335S}
{Shelyag}, S., {Sch{\"u}ssler}, M., {Solanki}, S.~K., {Berdyugina}, S.~V., \&
  {V{\"o}gler}, A. 2004, \aap, 427, 335

\bibitem[{{Shidei}(1936)}]{Shidei1936}
{Shidei}, T. 1936, Japanese J. Phys., 11, 23

\bibitem[{{Steimle} {et~al.}(1986){Steimle}, {Woodward}, \&
  {Brown}}]{Steimle1986}
{Steimle}, T.~C., {Woodward}, D.~R., \& {Brown}, J.~M. 1986, \jcp, 85, 1276

\bibitem[{{Tatum}(1966)}]{Tatum1966}
{Tatum}, J.~B. 1966, Publications of the Dominion Astrophysical Observatory
  Victoria, 13, 1

\bibitem[{{Tennyson} \& {Yurchenko}(2012)}]{Tennyson2012}
{Tennyson}, J. \& {Yurchenko}, S.~N. 2012, \mnras, 425, 21

\bibitem[{{Ubachs} {et~al.}(1986){Ubachs}, {Meyer}, {Ter Meulen}, \&
  {Dymanus}}]{Ubachs1986}
{Ubachs}, W., {Meyer}, G., {Ter Meulen}, J.~J., \& {Dymanus}, A. 1986, \jcp,
  84, 3032

\bibitem[{{van Dishoeck}(1987)}]{vanDishoeck1987}
{van Dishoeck}, E.~F. 1987, \jcp, 86, 196

\bibitem[{{V{\'a}zquez} {et~al.}(2007){V{\'a}zquez}, {Amero}, {Liebermann},
  {Buenker}, \& {Lefebvre-Brion}}]{Vazquez2007}
{V{\'a}zquez}, G.~J., {Amero}, J.~M., {Liebermann}, H.~P., {Buenker}, R.~J., \&
  {Lefebvre-Brion}, H. 2007, \jcp, 126, 164302

\bibitem[{{Watson}(2001)}]{Watson2001}
{Watson}, J.~K.~G. 2001, \apj, 555, 472

\bibitem[{{Zachwieja}(1995)}]{Zachwieja1995}
{Zachwieja}, M. 1995, Journal of Molecular Spectroscopy, 170, 285

\bibitem[{{Zachwieja}(1997)}]{Zachwieja1997}
{Zachwieja}, M. 1997, Journal of Molecular Spectroscopy, 182, 18

\end{thebibliography}
\onecolumn
\onltab{2}{
\begin{longtable}{lllllll}
\caption{Vacuum wavenumbers (cm$\rm^{-1}$) of clean CH A-X transitions in CS22942-019 used for the calculation of the constants. The numbers in parentheses are the observed - calculated value.}\label{tab:AX_lines}\\
\hline
 \multicolumn{7}{c}{           0 -           0 }\\
 J & R11ff & R11ee & Q11ef & Q11fe & P11ee & P11ff \\
 \hline
 27.5                    &            \ldots &            \ldots &            \ldots &            \ldots &    22713.66(  11) &           \ldots \\
 28.5                    &            \ldots &            \ldots &            \ldots &            \ldots &    22706.44(  23) &           \ldots \\
 29.5                    &    24138.80(  12) &            \ldots &            \ldots &    23400.12(  22) &            \ldots &           \ldots \\
 30.5                    &    24159.33(  22) &    24134.84(  41) &    23431.16(  13) &            \ldots &    22691.34(  19) &           \ldots \\
 31.5                    &            \ldots &    24151.75(  54) &    23437.20(  23) &            \ldots &    22683.31(  26) &   22708.81(   1) \\
\hline
\hline
\end{longtable}
}

\onltab{3}{
\begin{longtable}{lllllll}
\caption{Vacuum wavenumbers (cm$\rm^{-1}$) of clean CH B-X transitions in CS~22942-019 used for the calculation of the constants. The numbers in parentheses are the observed minus calculated value. The numbers in italics are predissociation lines where the accuracy of the position is low.}\label{tab:BX_lines}\\
\hline
 \multicolumn{7}{c}{           0 -           0 }\\
 J & R11ff & R11ee & Q11ef & Q11fe & P11ee & P11ff \\
 \hline
  0.5                       &               \ldots &               \ldots &               \ldots &               \ldots &               \ldots &              \ldots \\
  1.5                       &               \ldots &               \ldots &               \ldots &               \ldots &               \ldots &              \ldots \\
  2.5                       &               \ldots &               \ldots &               \ldots &               \ldots &               \ldots &              \ldots \\
  3.5                       &               \ldots &               \ldots &               \ldots &               \ldots &               \ldots &              \ldots \\
  4.5                       &               \ldots &               \ldots &               \ldots &               \ldots &               \ldots &              \ldots \\
\hline
\hline
\end{longtable}
}

\onltab{4}{
 \begin{table*}
\caption{Vacuum wavenumbers (cm$\rm^{-1}$) of clean CH C-X transitions in our stellar spectra used for the calculation of the constants. The numbers in parentheses are the observed minus calculated value.}\label{tab:CX_lines}
 \begin{tabular}{lllllll}
\hline
 \multicolumn{7}{c}{           2 -           2 }\\
 J & R11ff & R11ee & Q11ef & Q11fe & P11ee & P11ff \\
 \hline
  7.5                       &               \ldots &               \ldots &               \ldots &               \ldots &               \ldots &              \ldots \\
  8.5                       &               \ldots &               \ldots &               \ldots &               \ldots &      31192.72(  21) &              \ldots \\
  9.5                       &               \ldots &               \ldots &      31381.79(  24) &               \ldots &               \ldots &              \ldots \\
 10.5                       &               \ldots &               \ldots &      31369.38(  56) &               \ldots &      31121.54(  30) &              \ldots \\
 11.5                       &               \ldots &               \ldots &               \ldots &               \ldots &      31083.62(  21) &              \ldots \\
 12.5                       &               \ldots &               \ldots &               \ldots &               \ldots &      31043.85(   1) &              \ldots \\
\hline
\end{tabular}
\end{table*}
}

\onltab{5}{
 \begin{table*}
\caption{Vacuum wavenumbers (cm$\rm^{-1}$) of clean CH A-X transitions of $\rm ^{13}CH$ in our stellar spectra used for the calculation of the constants. The numbers in parentheses are the observed minus calculated value.}\label{tab:AX_C13H_lines}
\begin{tabular}{lllllll}
\hline
 \multicolumn{7}{c}{           1 -           1 }\\
 J                       & R11ff             & R11ee             & Q11ef             & Q11fe             &  P11ee            & P11ff \\
 \hline
 18.5                    &            \ldots &            \ldots &            \ldots &            \ldots &            \ldots &           \ldots \\
 19.5                    &    23758.86(  -3) &    23747.07(  -5) &            \ldots &            \ldots &            \ldots &           \ldots \\
 20.5                    &    23783.00( -11) &    23770.22(  -5) &            \ldots &            \ldots &            \ldots &           \ldots \\
 21.5                    &    23805.71( -21) &            \ldots &            \ldots &            \ldots &            \ldots &           \ldots \\
 22.5                    &    23826.87( -28) &            \ldots &            \ldots &            \ldots &            \ldots &           \ldots \\
 23.5                    &    23846.14( -42) &            \ldots &            \ldots &            \ldots &            \ldots &           \ldots \\
\hline
\hline
\end{tabular}
\end{table*}
}
\onltab{6}{
 \begin{table*}
\caption{Vacuum wavenumbers (cm$\rm^{-1}$) of clean CH B-X transitions of $\rm ^{13}CH$ in our stellar spectra used for the calculation of the constants. The numbers in parentheses are the observed minus calculated value. The numbers in italic are predissociation lines where the accuracy of the position is low.}\label{tab:BX_C13H_lines}
\begin{tabular}{lllllll}
\hline
 \multicolumn{7}{c}{           0 -           0 }\\
 J                       & R11ff             & R11ee             & Q11ef             & Q11fe             & P11ee             & P11ff \\
 \hline
 16.5                    &            \ldots &            \ldots &            \ldots &    25229.71(  -1) &            \ldots &       24873.91(   3) \\
 17.5                    &            \ldots &            \ldots &            \ldots &    25158.98(   7) &            \ldots &       24787.08(  -2) \\
 18.5                    &    25498.69(-384) &            \ldots &            \ldots &    25080.34(  -4) &            \ldots &       24693.96( -16) \\
 19.5                    &            \ldots &            \ldots &            \ldots &    24992.81(   1) &            \ldots &       24593.72( -26) \\
 20.5                    &            \ldots &            \ldots &            \ldots &    24894.23( -14) &            \ldots &       24485.60(  19) \\
 \hline
 J                       & R22ff             & R22ee             & Q22ef             &  Q22fe            & P22ee             & P22ff\\
\hline
\hline
\end{tabular}
\end{table*}
}
\onltab{7}{
 \begin{table*}
\caption{Vacuum wavenumbers (cm$\rm^{-1}$) of clean CH C-X transitions of $\rm ^{13}CH$ in our stellar spectra used for the calculation of the constants. The numbers in parentheses are the observed minus calculated value.}\label{tab:CX_C13H_lines}
\begin{tabular}{lllllll}
\hline
 \multicolumn{7}{c}{           0 -           0 }\\
 J                       & R11ff             & R11ee             & Q11ef             & Q11fe             & P11ee             & P11ff \\
 \hline
 15.5                    &            \ldots &    32223.03(   9) &            \ldots &            \ldots &            \ldots &           \ldots \\
 16.5                    &            \ldots &            \ldots &            \ldots &            \ldots &    31365.64( -25) &           \ldots \\
 17.5                    &            \ldots &    32266.81(  -2) &            \ldots &            \ldots &    31341.06( -23) &           \ldots \\
 18.5                    &            \ldots &            \ldots &            \ldots &            \ldots &            \ldots &           \ldots \\
 19.5                    &            \ldots &            \ldots &            \ldots &            \ldots &            \ldots &           \ldots \\
 20.5                    &            \ldots &            \ldots &            \ldots &            \ldots &    31266.86( -95) &           \ldots \\
\hline
\hline
\end{tabular}
\end{table*}
}

\end{document}